\documentclass[11pt]{article}
\usepackage[utf8]{inputenc}

\usepackage[paper=a4paper,margin=1in]{geometry}
\usepackage[header,title,page,titletoc]{appendix} 

\usepackage{amsmath}
\usepackage{bm}
\usepackage{amsfonts}
\usepackage{float}
\usepackage{braket}
\usepackage{graphicx}
\usepackage{subcaption}
\usepackage[none]{hyphenat} 

\newcommand\eea{\end{eqnarray}}
\newcommand\bea{\begin{eqnarray}}
\numberwithin{equation}{section}

\newcommand{\be}{\begin{equation}}
\newcommand{\ee}{\end{equation}}
\newcommand{\tr}{\text{tr} }
\def\({\left(}
\def\){\right)}
\def\[{\left[}
\def\]{\right]}

\newcommand{\al}{\alpha}

\usepackage[dvipsnames]{xcolor}
\definecolor{rossocorsa}{rgb}{0.83, 0.0, 0.0}
\definecolor{navyblue}{rgb}{0.0, 0.0, 0.5}

\usepackage{hyperref}
\hypersetup{
    colorlinks,
    citecolor=rossocorsa,
    filecolor=navyblue,
    linkcolor=navyblue,
    urlcolor=navyblue
}

\begin{document} 

\begin{titlepage}

\begin{center}

\phantom{ }

{\bf \LARGE{Holographic R\'enyi $n\to0$ entropy 
\vspace{.2cm}
\\and Euclidean fluids}}

\vskip 2cm

Cesar A. Ag\'on${}^{a}{}^{\dagger}$,  Horacio Casini${}^{b}{}^{*}$ and Pedro J. Martinez${}^{c}{}^{\ddagger}$

\vskip 1cm

${}^{a}$\small{  \textit{Institute for Theoretical Physics, Utrecht University}}

\vskip .15cm

\small{\textit{3584 CC Utrecht, The Netherlands}}

\vskip .3cm

${}^{b}$\small{  \textit{Instituto Balseiro, Centro At\'omico Bariloche}}

\vskip .15cm

\small{\textit{ 8400-S.C. de Bariloche, R\'io Negro, Argentina}}

\vskip .3cm

${}^{c}$\small{  \textit{Instituto de F\'isica La Plata, IFLP-CONICET}}

\vskip .15cm

\small{\textit{ Diagonal 113 e/ 63 y 64, La Plata Argentina}}

\vskip 2cm

\begin{abstract}{
We explore the holographic prescription for computing the refined R\'enyi entropies $\tilde S_n$ in the $n \to 0$ limit within the AdS$_{d+1}$/CFT$_d$ framework. This limit can be interpreted as a high-temperature regime with respect to the energy defined by the modular Hamiltonian of the state reduced to a subregion. To leading order in $n$, we find that the system attains local equilibrium and admits a CFT description in terms of a Euclidean, irrotational perfect fluid. This fluid exhibits vortex-like boundary conditions at the entangling surface. Guided by this physical picture, we construct an ansatz for the dual geometry in terms of the boundary fluid variables. We show that our anzats solves the Einstein's equations coupled to a cosmic brane at leading order in $n$, in agreement with Dong's proposal for the holographic dual to the refined R\'enyi entropy.  
The resulting conical singularity, signaling the brane's location, can be understood from this perspective as the bulk extension of the boundary fluid vortices. 

}\end{abstract}
\end{center}

\vspace{2cm}

\small{\vspace{2 cm}\noindent ${}^{\dagger}$cesaragon1@gmail.com \\\noindent 
${}^{*}$horaciocasini@gmail.com\\
${ }^{\ddagger}$martinezp@fisica.unlp.edu.ar
}

\end{titlepage}

\setcounter{tocdepth}{2}

{\parskip = .4\baselineskip \tableofcontents}
\newpage


\section{Introduction}
\label{Sec-uno}

For the last twenty years, holography has provided an important window to the exploration of quantum information measures in QFTs \cite{ryu2006holographic, Hung:2011nu, Lewkowycz:2013nqa,Rangamani:2016dms}. The geometrization provided by the Ryu-Takayanagi (RT) proposal for the von Neumann entropy \cite{ryu2006holographic} and the subsequent construction of geometric duals for other quantum information quantities such as R\'enyi entropy \cite{Hung:2011nu, dong2016gravity}, relative entropy \cite{Blanco:2013joa, Jafferis:2015del}, reflected entropy \cite{Dutta:2019gen}, and complexity \cite{Brown:2015bva}, among others \cite{Nakata:2020luh, Doi:2023zaf}, have proved fundamental in achieving a deeper understanding of these quantities in general QFTs. 

The refined R\'enyi entropy,\footnote{The refined R\'enyi entropy is related to the standard R\'enyi entropy $S_n$ via
\bea\label{ref-RE-RE}
\tilde{S}_n=n^2 \partial_n\left(\frac{n-1}{n} S_n\right)\,. \nonumber
\eea
In the $n\to0$, the R\'enyi entropy scales as $S_{n\to 0}\sim 1/n^{d-1}$, and the above relation simplifies to $\tilde{S}_{n\to 0}=d\, S_{n\to 0}$.} a one parameter generalization of the von Neumann entropy, is an interesting measure that provides a detailed characterization of the entanglement structure of the reduced state. In particular, it is known that via an inverse Laplace transform, one can reconstruct the eigenvalue distribution of the reduced state from this quantity. Given a reduced state\footnote{In this work, we denote as $\rho$ to the state reduced to a particular subregion $A$. Such state is typically denoted as $\rho_A$ but we removed the subindex $A$ to avoid cluttering.} $\rho$, the refined R\'enyi entropy is defined as 
\begin{equation}\label{Refined-Sn-Def}
\tilde S_n\equiv - n^2 \partial_n\left(\frac 1n \ln {\rm tr}\rho^n\right)\,.
\end{equation}  
State dependent quantities, such as the above, are notoriously hard to compute in QFTs. Most known computations of refined R\'enyi entropies involves free QFT states reduced to various geometries, and CFT states reduced to spherical regions.  

In holographic theories, the problem of computing the refined R\'enyi entropy becomes more tractable, although, explicitly worked out examples are scarce. As Dong proposed in \cite{dong2016gravity}, based on \cite{Lewkowycz:2013nqa}, the relevant dual geometry for the computation of $\tilde S_n$ is given by the solution to
the equations of motion of the Einstein's action coupled to a dynamical brane, i.e.
\begin{equation}\label{Grav+Brane-Action}
I=-\frac{1}{16 \pi G_N}\int\!\! d^{d+1}x\,\sqrt{|g|} \;\(R+\frac{d(d-1)}{L^2}\)+T_n\int_{\sim A} \!\!\!\!d^{d-1}x\, \sqrt{|h|}\;,\quad {\rm where}\quad T_n=\frac{n-1}{4 n G_N}\,,
\end{equation}
is the brane's tension, and $h$ the induced metric on the brane. In (\ref{Grav+Brane-Action}), the brane is homologous to the entangling region $A$, $L$ is the AdS radius, and $G_N$ is the appropriate Newton's constant. Dong's proposal states that the refined R\'enyi entropy in a holographic CFT,  can be computed as
\begin{equation}\label{Dong-prescription}
\tilde S_n=\frac{{\rm area(Cosmic Brane)}_n}{4G_N}\,,
\end{equation} 
where the area of the cosmic brane is evaluated on the solution to the coupled equations\footnote{One can add matter to \eqref{Grav+Brane-Action} and other modifications to model more generic boundary states \cite{dong2016gravity}, but we will not do so in this work.} from (\ref{Grav+Brane-Action}). Equation (\ref{Dong-prescription}) reduces to the RT minimal area prescription \cite{ryu2006holographic} in the brane's tensionless limit, i.e. $n\to 1$ limit.  Solving the non-linear differential equations of gravity plus brane is a formidable task and consequently, we have limited intuition about the general structure of ${\tilde S}_n$. Therefore, it is of interest to consider regimes in which the above setup simplifies allowing for analytic investigations of this fundamental quantity.

In this work, we study ${\tilde S}_n$ in the limit in which the R\'enyi parameter $n$ goes to zero. This limit captures information about the small eigenvalue distribution of the reduced state $\rho$ (equivalently, the large eigenvalue distribution of the modular Hamiltonian $K\equiv-\log \rho$, see e.g. \cite{Agon:2023tdi}). In holographic systems, the dual geometries simplify in this limit. The brane's tension becomes negative and divergent $T_n\to-\infty$, and a connection between the above gravitational problem and a problem in the theory of perfect fluids emerges.  The fluid problem consist in finding a (Euclidean) conformal relativistic irrotational perfect fluid in $d-$dimensional flat space with vortexes as boundary conditions. The vortexes are defects in the boundary theory sitting at the entangling surface.
This auxiliary fluid lives in the CFT, thus, its local temperature and fluid velocity are functions of the $d-$dimensional boundary coordinates.  We find that given a solution to the associated fluid equations of the above problem, we can construct the bulk geometry that solves the system \eqref{Grav+Brane-Action} at leading order in $n$, in the $n\to0$ limit. This geometry suffices to holographically compute $\tilde{S}_{n\to0}$ using \eqref{Dong-prescription}.

The irrotational condition means that given an inverse local temperature $\beta$ and fluid velocity $u_\mu$, satisfying the perfect fluid equations, we require them to additionally satisfy $d\wedge (u/\beta)=0$, and 
\begin{equation}\label{vortex-intro}
\oint_\Gamma \frac{u_\mu dx^\mu }{\beta}=1\;,
\end{equation}
where $\Gamma$ is any closed path encircling the entangling surface a single time. In section  \ref{Sec:d>2}, we constructed the bulk geometry (\ref{f3}) that solves (\ref{Grav+Brane-Action}) in the appropriate $n\to0$ limit
from a solution to the above set of fluid equations. Evaluating the refined R\'enyi entropy from this geometry results in
\begin{equation}\label{Us-result}
\tilde S_{n\to0}=\frac{{\rm area(Cosmic Brane)}_{n\to0}}{4G_N}=\frac{\sigma_s}{n^{d-1}} \int_A  \frac{d^{d-1}\vec{x}}{\beta(\vec{x})^{d-1}} +\dots 
\end{equation}
where $\sigma_s$ is the coefficient that appear in the relation between thermal entropy density and temperature, $s=\sigma_s T^{d-1}$ for the CFT at high temperatures. We use $\sigma_\varepsilon$ to denote the similar coefficient connecting energy density and temperature $\varepsilon= \sigma_\varepsilon T^d$. In holography, the coefficients are given by
\bea\label{stefan-B}
\sigma_{\varepsilon}=(d-1) \frac{L^{d-1}}{16  \pi G_N  }\(\frac{4\pi}{d}\)^d\,, \qquad\qquad \sigma_s=\frac{d }{d-1}\sigma_{\varepsilon}=\frac{L^{d-1}}{4 G_N  }\(\frac{4\pi}{d}\)^{d-1} \,,
\eea
and we will refer to them as Stefan Boltzmann constants.
The dots in \eqref{Us-result} are $O(n^{2-d})$ corrections, and the integral is over the  $d-1$ dimensional entangling region $A$ defined on the $\tau=0$ boundary hypersurface.

The fluid variables in (\ref{Us-result}) allows us to write down a manifestly covariant expression for ${\tilde S}_{n\to0}$ as the flux of a conserved boundary entropy current $J^\mu$ 
\bea\label{Us-result-2}
\tilde S_{n\to0} =\int_{\Sigma} \sqrt{|\gamma|}\,d^{d-1}\sigma\, \hat{n}_\mu J^\mu +\dots\qquad{\rm with } \qquad J^\mu=\frac{\sigma_s }{ n^{d-1}}\frac{u^\mu}{\beta(x)^{d-1}}\,,
\eea
where $\Sigma$ is an arbitrary boundary hypersurface homologous to the boundary entangling region $A$, with induced metric $\gamma_{ij}$. We denote by $\sigma$ the intrinsic coordinates parametrizing that surface, and $\hat{n}$ the unit vector normal to it. Equation (\ref{Us-result-2}) is the main result of this work. It relates the refined R\'enyi entropy in the ground state of a CFT reduced to a given subregion, with the thermodynamic entropy of a perfect fluid that circles the entangling surface and lives in the same space-time geometry. In this sense, holography acts as a tool to prove an equivalence between boundary quantities, but the final result no longer refers to any holographic object beyond the holographic Stefan-Boltzmann constant $\sigma_s$.

The bulk ansatz we propose as an approximate solution to \eqref{Grav+Brane-Action}, this is (\ref{f3}), is inspired by the black brane solution as well as the fluid/gravity correspondence \cite{Bhattacharyya:2007vjd, Bhattacharyya:2008xc, Hubeny:2010wp, Rangamani:2016dms} and can be motivated as follows. To calculate \eqref{Refined-Sn-Def} we need to compute the trace of the reduced density matrix $\rho$ raised to the power $n$, $\rho^n$ which has a path integral representation in an $n$-replicated geometry. Alternatively, we can compute $\tilde S_{n}$ as the von Neumann entropy of the density matrix $\tilde \rho_n:=\rho^n/\tr \rho^n$, by thinking of it as an excited state in an un-replicated geometry, this is 
\bea
\tilde S_{n}=-\tr \(\tilde \rho_n \log \tilde \rho_n\)\,.
\eea
The state $\tilde \rho_n$, can be interpreted as a thermal state with respect to the notion of energy provided by the modular Hamiltonian with temperature $1/n$, which diverges in the $n\to 0$ limit. At very high modular temperature one may expect these highly interacting systems to achieve local equilibrium and have a good hydrodynamic description. Notice that, despite the fact that the notions of energy provided by the modular Hamiltonian and the ordinary Hamiltonian are not the same, they are both proportional to each other for localized energetic excitations, as argued in \cite{Arias:2016nip,arias2017anisotropic,Agon:2023tdi}. This means that their associated temperatures will have a similar proportionality relation which support this intuition. Thus, in this regime, $\tilde \rho_n$ can be approximately described as a Euclidean fluid, whose thermodynamic entropy would agree with the von Neumann entropy of $\tilde \rho_n$. Such thermodynamic entropy can be computed holographically by constructing the dual geometry of such a fluid and computing its associated entropy \cite{Hubeny:2010wp}. This procedure is realized exactly in the case of a sphere/Rindler geometry as we will review in section \ref{HRE}, and generalized to arbitrary geometries, in the $n\to 0$ limit, in section \ref{Sec:d>2}.

A couple of comments on our approach are at hand. First, our result \eqref{Us-result-2} in combination with previous results from \cite{Agon:2023tdi} point towards a  more general description of $\tilde S_{n\to0}$ in terms of 
a local notion of temperature $\beta(\vec x)$ in this regime. In \cite{Agon:2023tdi}, and for asymptotically free theories, such notion of temperature was shown to emerge from the so called ``local temperatures'' $\beta_L(\vec x, \hat{p})$ \cite{Arias:2016nip,arias2017anisotropic}, where $\hat{p}$ marks an anisotropic dependence on the direction,  via some averaging procedure. In that case, we got an effective description governed by a Boltzmann equation which allows the construction of an infinite tower of conserved currents characterizing the system \cite{Agon:2023tdi}. For holographic CFT's we do not expect a Boltzmann description of collisionless free particles to arise due to strong interactions, and indeed, we instead land into the coarse grained description of a perfect fluid. Despite this fundamental difference, it is remarkable that in both free and holographic theories we obtain the same relation \eqref{Us-result-2} between $\tilde S_{n\to0}$ and a description in terms of a thermal gas or fluid respectively, supporting our claim of universality. Thus, we expect \eqref{Us-result-2} to hold also for non-holographic theories as well. 

Second, we stress that the equations of motion for a perfect fluid with vortex boundary conditions are by no means easy to solve and analytic solutions are out of reach for most cases. However, we also emphasize that a set of (Euclidean) perfect fluid equations in fixed $\mathbb{R}^{d}$ metric should be more tractable via numerical methods than the set of non linear equations of \eqref{Grav+Brane-Action}. This deserves future study. We are not attempting this numerical calculations in the present work.  

A third comment regards the use of the Fluid/Gravity approach \cite{Bhattacharyya:2007vjd, Bhattacharyya:2008xc} in the context of our work. The main difference with the standard real time approach is that we are describing a Euclidean fluid at the boundary and, as a consequence, we cannot use bulk lightcone coordinates. 
These coordinates were central in the development of the Fluid/Gravity program to find an algorithm capable to build a complete solution to the Einstein's equations order by order in a gradient expansion in terms of the dynamics of a boundary fluid \cite{Bhattacharyya:2007vjd}. As it stands, we lack the technology to correct our ansatz to the following order, but we stress that our solution is enough for the computation of $\tilde S_{n\to0}$ we are after.

The present paper is complementary to recent work on holographic R\'enyi entropies \cite{Dong:2023bfy,Penington:2024jmt} regarding corrections and comments to the original prescription made in \cite{dong2016gravity}. 
In these works a fixed area state basis is used to argue that, in the $n\to0$ limit, and for cases where there are multiple solutions for the brane, a different topology from the ones usually considered for holographic R\'enyi entropies should dominate. For a boundary system composed of two disconnected pieces, for example, they find that the bulk dominating saddle should be a manifold that does not have the topology of either the disconnected nor the connected case, which they denominate ``diagonal''. In the main text, we argue that our ansatz appears to be a reasonable limiting geometry in the $n\to0$ limit of such ``diagonal'' configurations.

On a similar note, our solutions for simple entangling surfaces are mathematically similar to the geometries found in \cite{Abajian:2023jye,Abajian:2023bqv} in the context of a different problem, namely, finding correlators of very heavy operators in the CFT. Our setup corresponds to their ``unphysical'' infinite negative mass regime. The calculation of R\'enyi entropy in the $n\to0$ limit forces us to focus on exactly the opposite regime as these works, and the solution has a different topology.

The paper is organized as follows. In section \ref{HRE}, we revisit the known solution to the holographic refined R\'enyi entropy, corresponding to the Rindler/sphere geometry as found in \cite{Hung:2011nu,dong2016gravity}. This solution is exact for all $n$. We reframe this solution in terms of the boundary fluid degrees of freedom to prepare the ground for our ansatz and to set-up notation. In section \ref{Sec:d>2} we present our ansatz and explain in detail in which sense it is an approximate solution for $d>2$ and $d=2$ separately. We also derive our prescription \eqref{Us-result}. 
In section \ref{Sec:d=2} we focus on the $d=2$ scenario from a CFT perspective and argue that we can infer the boundary stress tensor in the $n\to0$ limit. With this insight we explore the properties of its bulk dual and compare results with the previous section. 
In section \ref{Sec:Discussion} we give a short discussion and elaborate on future perspectives on our work. In appendix \ref{App:perfect-fluids} we review some properties of perfect fluids in the non-standard Euclidean signature for reference.


\section{Holographic refined R\'enyi Entropy \label{HRE}}

In this section we first review the well known case of the vacuum reduced to half-space or a sphere, where the solution of Dong's prescription for the R\'enyi entropies can be solved exactly for any $n$ and boundary dimension $d$. 
Part of the analysis can be generalized straightforwardly to other situations where there is a conformal Killing symmetry that leave invariant the region and the state. The analysis of the geometry in the $n\to 0$ limit suggests the construction of an anzats for $\tilde \rho_n$ valid for more general cases in this limit. 
The essential geometrical intuition will be that, at $n\to0$, the big negative tension $T_n\to-\infty$ ``flattens'' the brane. The picture is that of a mostly flat brane whose location varies very slowly, such that its longitudinal fluctuations are always heavily suppressed with respect to its inverse local temperature $\beta_n=n\,\beta\sim n\to0$ except very near to the entangling surface.
We will expand our exact solutions in the $n\to0$ limit and motivate our ansatz for more general entangling surfaces.

\subsection{Rindler space, spheres, and hyperbolic black holes}

In QFT, a prototypical region to study the physics of reduced systems is half space which is described by a Rindler geometry. The reason for this choice comes from the famous result by Bisognano and Wichmann \cite{Bisognano:1976za} which establishes that the vacuum modular Hamiltonian for half space is proportional to a boost symmetry generator of the theory. This result was generalized to ball regions in CFTs, using the known conformal map between half space and the ball \cite{Casini:2011kv,Hislop:1981uh}. Furthermore, a conformal map takes the vacuum state reduced to the ball shaped region of radius $R$ into a thermal state in a hyperbolic geometry on
\begin{eqnarray}\label{T0andR}
{\mathbb S}_{\beta_H}\times {\mathbb H}_R^{d-1}\;,\qquad\qquad \beta_H=2\pi R\,.
\end{eqnarray}
Through this relation, the thermal entropy in the above space directly yields the von Neumann entropy of the CFT ground state reduced to a spherical and/or Rindler regions. In the context of holography, such state corresponds to a hyperbolic black hole with inverse temperature $\beta_H$. Thus, holography further connects these quantities with the hyperbolic black hole entropy \cite{Casini:2011kv}.

  The set up described above was later extended to the computation of the R\'enyi entropies in \cite{Hung:2011nu}. We review this result following the prescription in \cite{dong2016gravity} in the picture in which there is no brane. In this picture, to compute $\tilde S_n$ we are instructed to replicate the boundary manifold $n$ times connecting the different copies across the entangling region, and look for vacuum solutions $B_n$ of pure gravity in the bulk that have the replicated manifold as boundary conditions. The refined R\'enyi entropy is computed as
\begin{equation}\label{Dong-Regular}
\tilde S_n=n^2\partial_n\(\frac{I_{\rm{reg}}[B_n]}{n}\)\,, \qquad I_{\rm{reg}}=-\frac{1}{16 \pi G_N}\int\!\! d^{d+1}x\,\sqrt{|g|} \;\(R+\frac{d(d-1)}{L^2}\)\,,
\end{equation}
where $I_{\rm{reg}}[B_n]$ is the gravity on shell action of the saddle solution $B_n$.

In the hyperbolic set up, the replicated boundary manifold is, c.f. \eqref{T0andR} 
\begin{eqnarray}\label{T0andR2}
{\mathbb S}_{\beta^H_n}\times {\mathbb H}_R^{d-1}\;,\qquad\qquad \beta^H_n\equiv n\, \beta_H\,.
\end{eqnarray}
A vacuum solution to Einstein gravity with the boundary \eqref{T0andR2} can be found as an Euclidean Hyperbolic black hole, given by  \cite{Hung:2011nu}
\begin{equation}\label{Rindler-Myers-metric}
ds_{BH}^2=L^2 g(\rho)\frac{d\tau_H^2}{R^2}+\frac{d\rho^2}{g(\rho)}+\rho^2 d{\cal H}^2_{d-1}\,, \qquad g(\rho)=\frac{\rho^2}{L^2}-\frac{\rho_h^{d-2}}{\rho^{d-2}} \( \frac{\rho_h^2}{L^2}-1\)-1\,,
\end{equation}
where $d{\cal H}_{d-1}$ is the line element of ${\mathbb H}^{d-1}$ of unit radius and
$\tau_H \in [0,\beta^H_n)$, $ \rho\in[\rho_h,\rho_c]\,.$ 
The replicated boundary manifold \eqref{T0andR2} is recovered at $\rho=\rho_c$ in the $\rho_c \to \infty$ limit via 
\begin{eqnarray}\label{bound-metric-BHH}
ds_{CFT}^2\equiv\lim_{\rho_c\to \infty}\frac{R^2}{\rho_c^2}\;ds^2_{BH}|_{\rho=\rho_c}=d\tau_H^2+R^2 d{\cal H}^2_{d-1}\,.
\end{eqnarray}
The regularity of the solution is guaranteed by fixing $\rho_h$ so that near $\rho\sim \rho_h$ the euclidean horizon behaves like a regular disk. We can do this by defining a radial coordinate $\zeta$ as
\begin{eqnarray}
d\zeta^2=\frac{d\rho^2}{g(\rho)}\sim \frac{d\rho^2}{g'(\rho_h)(\rho-\rho_h)}\,,  \qquad
\qquad g(\rho)\sim \frac{g'(\rho_h)^2}{4}\zeta^2\,,\qquad (\,\rho\sim\rho_h\,)\,,
\end{eqnarray}
which leads to the following near horizon geometry
\begin{equation}\label{near-hor-hyper}
ds_{BH}^2\sim  d\zeta^2+ \zeta^2\( L\frac{g'(\rho_h)}{2} \frac{d\tau_H}{R}\)^2 + \rho_h^2 \; d{\cal H}^2_{d-1}\,,\qquad (\,\rho\sim \rho_h\,)\,.
\end{equation}
The regularity condition can be imposed by defining $\rho_h$ as 
\begin{equation}\label{rho-h-n}
L\frac{g'(\rho_h)}{2} 
\oint\frac{d\tau_H}{R}
=L\frac{g'(\rho_h)}{2} \frac{\beta^H_n}{R}
=2\pi
\quad\Rightarrow\quad
\rho_h\equiv\rho_{h;n}=\frac{L}{nd}  \left(1+\sqrt{1+d(d-2) n^2}\right)\,,
\end{equation}
which is equivalent to ask that the term within parenthesis in \eqref{near-hor-hyper} corresponds to an angular coordinate with period $2\pi$.
A direct computation of \eqref{Dong-Regular} using \eqref{Rindler-Myers-metric} yields the well-known result \cite{Hung:2011nu}
\begin{equation}\label{tildeSn-HYP}
\tilde{S}_n=\frac{\rho^{d-1}_{h;n}}{4G_N} {\rm vol}({\mathbb H}^{d-1})\,.
\end{equation}
An alternative picture of the same prescription is given in \cite{dong2016gravity} as follows. The $B_n$ solution above does not spontaneously break the $\mathbb{Z}_n$ symmetry induced by the replica trick at the boundary. Then one can consider a quotiented manifold $\hat B_n= B_n/\mathbb{Z}_n$ which is locally identical to $B_n$ except for a codimension-2 singularity sitting at the fixed points of the $\mathbb{Z}_n$ replica symmetry, inducing a deficit angle
\begin{eqnarray}\label{con-def-OK}
\Delta \theta=2\pi\(\frac{n-1}{n}\)\,.
\end{eqnarray}
The resulting manifold $\hat B_n$ can then be seen to be a solution to the equations of motion of \eqref{Grav+Brane-Action}, where the backreacting brane with tension $T_n$ puts the correct deficit angle at the original fixed points of the $\mathbb{Z}_n$ symmetry in $B_n$. In this picture the refined R\'enyi entropy is computed as \eqref{Dong-prescription}. Notice that the topology of the boundary of $\hat B_n$ is now the one of the original system, but one can check that the state described holographically by $\hat B_n$ is not the vacuum for $n\neq1$. Notice also that in this ``brane picture'', the parameter $n$ enters only as the tension of the brane $T_n\in\mathbb{R}$, so that the analytic extension often required for computing $\tilde S_n$ is built in the solution $\hat B_n$ provided one can solve the system \eqref{Grav+Brane-Action}. From this point forward in our work, we will always consider Dong's prescription in this so-called ``brane picture''.

For our current example of a system in \eqref{T0andR}, the solution $\hat B_n$ can be easily found from $B_n$ in eq. \eqref{Rindler-Myers-metric} since the brane is sitting at the horizon, so that locally they both look the same with $\rho_h=\rho_{h;n}$, but now $\tau_H\in[0,\beta_H)$ which matches the size of the system in \eqref{T0andR} and induces the correct conical deficit \eqref{con-def-OK}. This picture is fundamental in our interpretation of the boundary state $\tilde{\rho}_n$ as being an excited state living in a single copy of the boundary geometry. Thus, in this perspective the geometry of the back-reacted brane describes the dual geometry of that excited state whose von Neumann entropy is given by the area of the associated brane. Using \eqref{Dong-prescription} on $\hat B_n$ recovers \eqref{tildeSn-HYP}.

\subsection{A perfect fluid stress-energy tensor }

We now interpret the state dual to the brane as the excited state $\tilde \rho_n$ \eqref{T0andR} and study its properties via holography. 
We can check that the back-reacted geometry induces a nontrivial stress tensor that vanishes at $n=1$ or equivallently $\rho_{h;n}=L$, see \eqref{rho-h-n}, but will otherwise be non-zero. One can read this information from the near boundary expansion of $\hat B_n$, following either the standard Fefferman-Graham procedure \cite{Fefferman:95, deHaro:2000vlm} or extracting the result from the bulk Brown-York tensor \cite{Balasubramanian:1999re}. The result reads, recall $\beta_H=2\pi R$, 
\begin{equation}\label{stress-hyper}
\begin{aligned}
T^{H}_{\mu \nu}
= \frac{L^{d-1} }{16  \pi G_N  }\frac{\rho_{h;n}^{d-2}}{L^{d-2}}\left(\frac{\rho_{h;n}^2}{L^2}-1\right)\frac{(2\pi)^d}{\beta_H^d}\(g^H_{\mu \nu}-d \,u^H_\mu u^H_\nu\)\,, 
\end{aligned}
\end{equation}
where $g^H_{\mu \nu}$ is the metric on the boundary hyperbolic geometry ${\mathbb S}_\beta\times {\mathbb H}^{d-1}_R$, and $u_\mu^H$ is a unit vector $u_H^2=1$ with respect to $g^H_{\mu \nu}$ that points along the Euclidean time direction $u_H^\mu\partial^H_\mu=\partial_{\tau_H}$. 
The stress energy tensor in (\ref{stress-hyper}), is conserved, traceless and has the form of a perfect fluid. This is
\be\label{EPFluidsT-bulk}
T_{\mu\nu}=P\(g^H_{\mu \nu}-d \,u^H_\mu u^H_\nu\) \,, \qquad{\rm with}  \qquad \nabla^{\mu}T_{\mu\nu}=0,  \qquad {\rm and}  \qquad  T^\mu_{\,\,\,\, \mu}=0 \,,
\ee
where the energy density $\varepsilon:=-u^\mu u^\nu T_{\mu\nu}=(d-1)P$ is given by\footnote{Notice that $\varepsilon>0$ as $n<1$ and viceversa. This might seem counter-intuitive at first but recall from \eqref{rho-h-n} that $n\to0$ represents a large mass black hole, $\rho_h\gg L$ which is a solution with positive energy density. The $n>1$, describes a state with negative energy, and in some cases  the associated solution can become unstable  \cite{Belin:2013dva}.  }  
\begin{eqnarray}
\varepsilon=\frac{\sigma_n}{ \beta_H^{d}} \,, \qquad {\rm with} \qquad \sigma_n=(d-1)(2\pi)^d \frac{L^{d-1}}{16  \pi G_N  }\frac{\rho_{h;n}^{d-2}}{L^{d-2}}\left(\frac{\rho_{h;n}^2}{L^2}-1\right)\;.
\end{eqnarray}
In the $n\to0$  limit $\sigma_n\approx \sigma_{\varepsilon}/n^{d}$, and the above expressions simplify to 
\be\label{energy-density}
\varepsilon\sim \frac{\sigma_{\varepsilon}}{n^d\beta_H^d}=\frac{\sigma_{\varepsilon}}{(\beta^H_n)^d}\,,\qquad \sigma_{\varepsilon}=(d-1) \frac{L^{d-1}}{16  \pi G_N  }\(\frac{4\pi}{d}\)^d,\qquad (\,n\to0\,)
\ee
where $\sigma_{\varepsilon}$ is the Stefan-Boltzmann constant relating energy density and temperature in flat space for the holographic CFT as defined around \eqref{stefan-B}. This same constant can be read off from the geometry of an ordinary black brane. Notice that $n$ now appears only alongside $\beta_H$ in the combination $\beta_n^H \equiv n\,\beta_H$ which makes manifest our interpretation of $\tilde{\rho}_n$ as a high temperature thermal state in the $n\to 0$ limit.

A perfect fluid like the one described by \eqref{EPFluidsT-bulk} has an associated entropy current $J_s^\mu$. At $n\to 0$ this current is given by 
\begin{equation}
J_s^\mu\equiv \sigma_s\frac{u_H^{\mu}}{(\beta^H_n)^{d-1}}\,,\qquad\qquad{\rm with }\qquad\qquad \nabla_{\mu}J_s^{\mu}=0\,,
\end{equation}
which is conserved by the equations of motion of the perfect fluid, see Appendix \ref{App:perfect-fluids}. The constant $\sigma_s$ is given by \eqref{stefan-B}.  This normalization, makes the flux of $J^\mu$ across a surface homologous to the entangling region, i.e. ${\mathbb H}^{d-1}_R$, be equal to $\tilde S_{n\to0}$, see \eqref{rho-h-n}, i.e. 
\begin{equation}\label{S0-Rindler}
\tilde S_{n\to0}
=\int_\Sigma d\Sigma\, \hat{n}_\mu J_s^\mu
=\sigma_s \frac{{\rm vol}({\mathbb H}_R^{d-1})}{(\beta^H_n)^{d-1}}\,,
\end{equation}
where $\Sigma$ is any surface homologous to ${\mathbb H}_R^{d-1}$ inside ${\mathbb H}_R^{d-1}\times S_\beta$, and $\hat{n}$ its unit normal vector. 

We can transform this stress tensor and currents back into the flat Euclidean geometry, using either the Euclidean map that takes the hyperbolic space to the ball region or the one that takes it to half space. Below, we call $\Omega(x)$ the conformal factor that turns the hyperbolic metric into ${\mathbb R}^d$ after any of these change of coordinates. In either case we would obtain the expectation value of the stress-energy tensor characterizing the excited state $\tilde \rho$ for the associated region. Using this, one gets
\begin{eqnarray}\label{Conf-Transf}
T^{{\mathbb R}^d}_{\alpha \beta}=\Omega^{d-2}\frac{\partial x_H^\mu}{\partial x^\alpha}\frac{\partial x_H^\nu}{\partial x^\beta}T^{H}_{\mu \nu}=\frac{L^{d-1} }{16  \pi G_N  }\frac{\rho_{h;n}^{d-2}}{L^{d-2}}\left(\frac{\rho_{h;n}^2}{L^2}-1\right)\frac{(2\pi)^d}{\beta(x)^d}\Big(\delta_{\alpha \beta}-d \,u_\alpha(x) u_\beta(x)\Big)\,.
\end{eqnarray}
The structure of the stress-energy tensor in the flat geometry will still be the one of a perfect fluid, but as a result of the transformation we have obtained a space-dependent notion of temperature from the conformal factor $\Omega(x)$. Namely, the new notion of temperature and fluid velocity will be 
\begin{equation}\label{beta-Omega}
\beta(x)\equiv \frac{\beta_H}{\Omega(x)}\,,
\qquad\qquad  u_\mu(x)=\frac{1}
 {\Omega(x)}\,\frac{\partial x_H^\nu}{\partial x^\mu} \, u^H_\nu\,.
\end{equation}
To be concrete, we get for the temperatures 
\begin{eqnarray}\label{betaSR}
\beta_{S}(x)= \pi R  \sqrt{1+2\frac{( \tau_S^2- r_S^2)}{R^2}+\frac{( r_S^2+ \tau_S ^2)^2}{R^4}}\;,\quad {\rm and}\quad \beta_{R}(x)=2\pi r\,,
\end{eqnarray}
where subindex $S$ and $R$ stands for sphere and Rindler respectively. Here $r_S$ is a radial coordinate centered at the origin in the case of the ball geometry, $\tau_S$ is the non-compact Euclidean time coordinate after the map to the sphere, and $r$ is the Euclidean distance from the point to the Rindler wedge, including the Euclidean time component. Similarly, the fluid velocities are
\begin{align}\label{uSR}
u^\mu_S \partial_\mu = \frac{\(R^2+\tau_S^2-r_S^2\)\partial_{\tau_S}+2r_S \tau_S\,\partial_{r_S} }{\sqrt{R^4+2R^2(\tau_S^2-r_S^2)+(r_S^2+\tau_S^2)^2}}\,,\quad{\rm and}\quad u^\mu_R\partial_\mu
=\frac{1}{r}\partial_{\theta}\,.
\end{align}
where $\theta$ is the angular coordinate on the plane orthogonal to the Rindler wedge. 

We can also revisit the conical deficit around the brane in these fluid degrees of freedom now rendered local on the plane.
It is straightforward to show that 
\begin{eqnarray}\label{Circulation-HBH}
\int_0^{\beta_H} \frac{d\tau_H}{2\pi R}=\oint \frac{u^H_\mu dx_H^\mu}{\beta_H}=\oint \frac{u_\mu dx^\mu}{\beta}\equiv \oint \omega_\mu dx^\mu =1\,.
\end{eqnarray}
where recall here that $\tau_H\in[0,\beta_H)$ as in the original un-replicated system \eqref{T0andR}, since we are working in the ``brane picture''.
The closed path in this equation is any path that links once the entangling surface. This follows from the fact that the vector $ \frac{u^\mu}{\beta}=\frac{\partial x_H^\mu}{\partial x^\nu} \, u_H^\nu$ is the gradient of $\tau_H$, and its curl vanishes. For a perfect fluid, the quantity $\frac{u^\mu}{\beta}\equiv \omega_\mu$ is the relativistic vorticity tensor, see App. \ref{App:perfect-fluids}.

\subsection{The hyperbolic black hole in fluid variables }

In what follows, we will change variables so that the metric \eqref{Rindler-Myers-metric} more naturally incorporates the fluid variables. This will help to motivate the general ansatz we obtain in the $n\to0$ limit.
We begin by changing both the bulk coordinate as well as the cut-off surface to automatically incorporate the conformal transformations to flat space in \eqref{Conf-Transf}, i.e.
\begin{equation}\label{z-var-domain}
z= L\,\frac{  \rho_{h;n}}{\rho} \;,\qquad z\in \left[  2\pi\epsilon\frac{ \rho_{h;n} }{\beta}  \,,L\right]\;.
\end{equation}
where $\epsilon\ll L$ is the cutoff,
to obtain
\begin{equation}\label{z-pre-Rindler-metric}
ds^2=\frac {L^2 \rho_{h;n}^2}{z^2 R^2}\Big[h(z)d\tau_H^2+ R^2d{\cal H}^2_{d-1}\Big] +\frac{L^2}{z^2}\frac{dz^2}{h(z)}\,, \qquad h(z)=1-\frac{z^d}{L^d}\(1-\frac{L^2}{\rho_{h;n}^2}\)-\frac{z^2}{\rho_{h;n}^2}\,.
\end{equation}
The second step is to identify the fluid variables in the metric. For this goal, it is convenient to use relations (\ref{beta-Omega}), and (\ref{Circulation-HBH}) as well as the conformal transformation from Hyperbolic to flat space. This is 
\bea
g^H_{\mu \nu}dx_H^\mu dx_H^\nu:=d\tau_H^2+R^2 d{\cal H}^2_{d-1}=\Omega^2\[(u_\mu dx^\mu)^2+ (\delta_{\mu \nu} -u_\mu u_\nu)dx^\mu dx^\nu\]\,,
\eea
where the term in brackets is the ordinary length element in Euclidean space, with the projection along the fluid velocity $u_\mu$ explicitly separated.  The conformal factor $\Omega$ can be expressed in terms of the fluid variables in flat space as $\Omega= 2\pi R/\beta $. The above separation is convenient, since $d\tau_H=u_\mu^H dx_H^\mu=\Omega\, u_\mu dx^\mu$ and thus, we can identify each piece of the metric independently as
\begin{equation}
d\tau_H^2=(2\pi R)^2 \frac{(u_\mu dx^\mu)^2}{\beta^2}\qquad{\rm and}\qquad R^2\,d{\cal H}^2_{d-1}=(2\pi R)^2\frac{dx_\mu dx^\mu-(u_{\mu}dx^\mu)^2}{\beta^2}\,.
\end{equation}
Plugging these expressions into \eqref{z-pre-Rindler-metric} leads to  
\begin{equation}\label{Rindler-fluid-metric-Brane}
ds^2=\frac {(2\pi)^2L^2 \rho_{h;n}^2}{z^2 \beta^2}\Big[h(z)(u_\mu dx^\mu)^2 + \left(dx_\mu dx^\mu-(u_{\mu}dx^\mu)^2\right)\Big] +\frac{L^2}{z^2}\frac{dz^2}{h(z)}\,.
\end{equation}
This particular form of the metric is especially useful to explore the region near the brane, since it is defined at constant $z=L$, see \eqref{z-var-domain}. To reach the boundary it is convenient to change variables to $\zeta =z \,\beta/(2\pi \rho_{h;n})$ leading to
\begin{equation}\label{Rindler-fluid-metric-Bound}
ds^2=\frac {L^2 }{\zeta^2}\Big[\tilde h(\zeta)(u_\mu dx^\mu)^2 + \left(dx_\mu dx^\mu-(u_{\mu}dx^\mu)^2\right)\Big] +L^2 \frac{\left[d\ln\(\zeta/\beta\)\right]^2}{\tilde h(\zeta)}\,, \quad \zeta\in\left[\epsilon,\frac{L \beta}{2\pi\rho_{h;n}}\right]\;,
\end{equation}
where we have introduced $\tilde h(\zeta)\equiv h\left(\frac{2\pi \rho_{h;n}}{\beta} \zeta\right)$ to shorten notation.
Notice that in the coordinates \eqref{Rindler-fluid-metric-Bound} the metric now includes off diagonal terms 
$$ d\zeta d\beta=\partial_\mu \beta d\zeta dx^\mu \qquad\text{ as well as } \qquad d\beta^2=\partial_\mu \beta \partial_\nu \beta dx^\mu dx^\nu$$ 
hidden in the last term by the notation. The boundary metric is recovered as 
\begin{equation}\label{fluid-flat-bound}
ds_{CFT}^2\equiv\lim_{\epsilon\to 0}\frac{\epsilon^2}{L^2}\;ds^2|_{\zeta=\epsilon}= dx_\mu dx^\mu\,,
\end{equation}
where we have used that $\tilde h(\zeta)\to1$ and that the off diagonal components $d\zeta d\beta$ and $d\beta^2$ are subleading as $\zeta\to0$. 
The metrics \eqref{Rindler-fluid-metric-Brane}-\eqref{fluid-flat-bound} are  crucial for the general ansatz we propose in the next section. This shows how the perfect fluid variables enter the metric in this example for which we know the exact bulk solution. We have also managed to obtain an asymptotic flat metric so that the perfect fluid lives in flat space. Notice that in \eqref{Rindler-fluid-metric-Bound}, the bulk cut-off is constant $\zeta=\epsilon$ but the picture is that of a space-dependent brane, the introduction of off diagonal components to the metric was crucial to prevent singularities to develop near the brane. 

\subsubsection{Testing the \texorpdfstring{$n\to0$}{n->0} limit on an exact solution }\label{Testing n0}


We now consider an $n\to0$ expansion of our exact solutions \eqref{Rindler-fluid-metric-Brane}-\eqref{Rindler-fluid-metric-Bound}. The expansion of \eqref{Rindler-fluid-metric-Brane} to leading order in $n\to0$ is straightforward. The important pieces are,
\begin{equation}
\rho_{h;n}\sim \frac{2}{nd}\Big(1+O(n^2)\Big)\,, \qquad h(z)
\sim 1-\frac{z^d}{L^d}+O(n^2)\,,  \qquad z\in \left[ \frac{ 4\pi }{\beta_n d} \epsilon  \,,L\right]\;.
\end{equation}
Thus, we can approximate \eqref{Rindler-fluid-metric-Bound} to leading order as 
\begin{equation}\label{Rindler-order-1}
ds^2\sim\frac {(4\pi L)^2}{z^2 \beta_n^2 d^2}\Big[h_0(z)(u_\mu dx^\mu)^2 + \left(dx_\mu dx^\mu-(u_{\mu}dx^\mu)^2\right)\Big] +\frac{L^2}{z^2}\frac{dz^2}{h_0(z)}\,, \qquad z\in \left[ \frac{ 4\pi }{\beta_n d} \epsilon  \,,L\right]\;.
\end{equation}
where we have defined $h_0(z):=1-z^d/L^d$ for shorthand. One can check by direct computation that $\tilde S_{n\to0}$ obtained from the metric above reproduces \eqref{S0-Rindler}. Notice also that \eqref{Circulation-HBH} ensures that the brane has the correct tension throughout. We should stress that \eqref{Rindler-order-1} is not a solution to the system \eqref{Grav+Brane-Action} but only an approximate solution to leading order in $n$ that keeps enough information to compute $\tilde S_{n\to0}$ exactly. 

A number of comments to \eqref{Rindler-order-1} are at hand. First, notice that the $n$ dependence has simplified drastically in this limit and enters only in $\beta_n=n \,\beta$ which is vanishing as $n\to0$. Put in other terms, this limit can be regarded as a high temperature limit $\beta_n\to0$ of the boundary fluid, as we anticipated in \eqref{energy-density}. 
Taking $\{\beta_n,u_\mu\}$ as the fluid variables clarifies the nature of our expansion. Consider the picture of the sphere were the we have an explicit scale $R$ to compare fluctuations against $\beta^S_n$. One can readily check by explicit computation that for any point of the system, see \eqref{betaSR} and \eqref{uSR}
\begin{equation}
\partial_{\mu}\ln\beta^S_n\sim R^{-1}\,, \qquad\qquad \partial_{\mu}u_S^\nu\sim R^{-1}\,, \qquad\qquad \beta^S_n\sim n R\,,
\end{equation}
i.e. that the rate of change in the longitudinal directions of a fluid in a system of size $R$ are of order $R^{-1}$. We then can check that
\begin{equation}\label{grad-param}
\beta^S_n \partial_{\mu}\ln\beta^S_n\ll 1\,, \qquad\qquad \beta^S_n \partial_{\mu}u_S^\nu\ll1\,,\qquad\qquad \text{ as }\; n\to0\,,
\end{equation}
which enables a fluid gradient expansion in the boundary stress tensor \eqref{Conf-Transf} in the form \eqref{grad-T-1} as explained in App. \ref{App:perfect-fluids}.
Now, the standard gradient expansion used in Fluid/Gravity correspondence is made starting from a point $p$ in the boundary and approximating the global bulk solution via tube-wise solutions into the bulk around each boundary point $p$. These solutions solve exactly the bulk equations in the direction of the bulk, but perturbatively order by order in $\beta^S_n/R$. The expansion can be shown to be valid in a radius of $(\beta^S_n/R)^{-1}$ around $p$. In the Fluid/Gravity correspondence, $\beta^S_n/R$ is taken to be small but finite and a physical parameter of the system. 
In our limit of interest, $\beta^S_n/R\sim n\to0$, 
$n$ controls the gradients. This allowed us to turn the gradient expansion into a more standard series expansion in $n$ and find a global solution up to order $n^0$ corrections.

Finally, notice also from \eqref{Rindler-order-1} that an order of limits $\epsilon\ll n \ll1$ is mandated for the geometric picture to make sense, otherwise the boundary would touch the brane sitting at $L$ at some point, since both $\epsilon\ll1$ and $n\ll1$. In these coordinates, for any point at which $\beta\neq0$, since $\epsilon/\beta_n\to0$, we get a global picture of a flat black brane sitting at $z=L$ with $z\sim[0,L]$ with ``spikes'' that reach the boundary at the points where $\beta=0$. Alternatively, taking the $n\to0$ limit of the metric \eqref{Rindler-fluid-metric-Bound}, as $\epsilon\ll n \ll1$ one gets a picture of a moving brane very close to a flat boundary. 

\section{Holographic dual of small \texorpdfstring{$n$}{n} R\'enyi entropies \label{Sec:d>2}}

In the previous section we studied in detail the refined R\'enyi entropy of the ground state of a holographic CFT reduced onto a spherical and/or Rindler region. We analyzed  the dual backreacted brane geometry both for finite $n$ as well as in the $n\to 0$ limit. In this section, we will extrapolate some of the features found in this specific example and propose an ansatz for the brane geometry dual to $\tilde{\rho}_n$ which explicitly incorporate such features. In summary, our ansatz part from the idea that the expectation value of the boundary stress-energy tensor in the aforementioned state, as $n\to0$, takes the form of a vorticity free perfect fluid for arbitrary regions with a delta function vorticity source located at the boundary of the entangling region.
\subsection{Ansatz for geometric dual to \texorpdfstring{$\rho^n/{\rm tr} \rho^n$}{redrho}}\label{BlackBraneAprox}
We would like to start with the assumption that the state $\tilde{\rho}_n:=\rho^n/\tr \rho^n$ in the $n\to 0$  limit is well approximated by an Euclidean fluid at local equilibrium. With the fluid variables we will construct the bulk ansatz and check it satisfies the equations \eqref{Grav+Brane-Action} at the leading order in $n$. 

  A canonical example of an asymptotically AdS spacetime that describes a thermal boundary state (whose stress tensor has the form of a perfect fluid) is the black brane solution:
\begin{equation}\label{black brane}
\begin{aligned}
ds^2=\frac{L^2}{z^2} \left(f(z)\, d\tau^2+ d\vec{x}^2 \right)+ \frac{L^2}{f(z)}\frac{dz^2}{z^2} \,,\qquad z\in[\epsilon,z_h]\,, 
\end{aligned}
\end{equation}
where $\tau\in [0, \beta]$ is a periodic Euclidean time coordinate whose period equal the boundary temperature and $f(z)=1-(z/z_h)^d$ is the blackening factor of the brane. The associated boundary fluid, is homogenous, with constant inverse temperature $\beta=4\pi z_h/d$, and velocity $u^\mu \partial_\mu =\partial_\tau$. The associated boundary stress energy tensor can be obtained by the usual near boundary Feffermann Graham expansion \cite{Fefferman:95, deHaro:2000vlm} 
\bea\label{bdy-SET-brane}
 T_{\mu \nu} =\frac{L^{d-1}}{16\pi G_N}\frac{1}{z_h^d}\(\delta_{\mu \nu}-u_\mu u_\nu\)=\frac{L^{d-1}}{16\pi G_N}\(\frac{4\pi }{d}\)^d\frac{1}{\beta^d}\(\delta_{\mu \nu}-u_\mu u_\nu\)\,,
\eea
in agreement with \eqref{EPFluidsT-bulk} and  (\ref{energy-density}).

Motivated by this solution, the solution (\ref{Rindler-fluid-metric-Bound}) and the fluid-gravity correspondence, we propose the following ansatz:
\begin{equation}\label{f3}
\begin{aligned}
ds^2=\frac{L^2}{z^2} \left(f(z,x)\, (dx\cdot u)^2+ (dx^2-  (dx\cdot u)^2) \right)+ \frac{L^2}{f(z,x)}\left(\frac{dz}{z}-\frac{dz_h}{z_h}\right)^2 \,,\qquad z\in[\epsilon,z_h(x)]\,. 
\end{aligned}
\end{equation}
This differs from (\ref{black brane}) in that we have promoted $u$ and $z_h$ to be functions of the boundary coordinates, $f(z,x)=1-(z/z_h(x))^d$. 
The modification in the last term of the metric guarantees that the factor $1/f(z,x)$ does not blow up at the horizon surface $z=z_h(x)$ while preserving the AdS asymptotics. This ansatz has the same structure than (\ref{Rindler-fluid-metric-Bound}), and the function $\tilde{h}(z/z_h(x))\to  f(z,x)$ in the $n\to 0$ limit.

At leading, large temperature $\beta(x)^{-1}$, the boundary stress tensor associated to this geometry is equal to (\ref{bdy-SET-brane}) but now with position dependent velocity $u=u(x)$ and inverse temperature $\beta(x)\propto z_h(x)$. This is easy to argue, since any correction to (\ref{bdy-SET-brane}) should involve derivatives of either $u_\mu$ or $\beta$, but from dimensional analysis such derivative terms must necessarily come with a respective factor of $\beta$, and therefore, it will necessarily be suppressed as $\beta \to 0$ with respect to the leading perfect fluid result. Such corrections can be explicitly computed via the Brown York tensor \cite{Balasubramanian:1999re}, or the Fefferman Graham expansion  \cite{Fefferman:95, deHaro:2000vlm}.

 The above metric has been designed to incorporate explicitly the correct conical singularity at the horizon, which in our case corresponds to the location of the back-reacted brane. To accomplish this, it is convenient to perform the following bulk rescaling $\zeta= z/z_h(x)$ into an adimensional bulk coordinte $\zeta$ which leads to the metric
\begin{equation}
\label{ansatz-2}
ds^2=\frac{L^2}{\zeta^2} \left[f(\zeta)\, \left(\frac{dx\cdot u}{z_h(x)}\right)^2+ \left(\frac{dx^2-  (dx\cdot u)^2}{z_h(x)^2}\right) \right]+ \frac{L^2}{f(\zeta)}\frac{d\zeta^2}{\zeta^2}\,,\qquad \zeta\in\left[\frac{\epsilon}{z_h(x)},1\right] \,.
\end{equation}
This transformation flattened the location of the horizon and give rise to the standard black brane geometry (\ref{black brane}). The standard analysis performed in (\ref{near-hor-hyper}), (\ref{rho-h-n}) to find a smooth geometry can be carried out here to make the near horizon geometry have a conical deficit singularity $\Delta \theta=2\pi(1-1/n)$ as in \eqref{con-def-OK}.  This requires the following condition on the boundary data 
\begin{equation}
   \frac{d}{4\pi } \oint_\Gamma \frac{dx\cdot u}{z_h(x)} =\frac{ 1}{n}\,,
\end{equation}
where the curve $\Gamma$ circles the boundary of the entangling surface $\partial A$ once.

The above condition can be interpreted as a vorticity condition on the boundary fluid via the identification $z_h(x):=d \beta_n(x)/4\pi= nd \beta(x)/4\pi$ with the precise source term
\begin{eqnarray}\label{circulation-source}
\partial_\mu w_\nu-\partial_\nu w_\mu= (\hat{n}^1_\mu \hat{n}^2_\nu-\hat{n}^1_\nu \hat{n}^2_\mu)\delta^{2}\({x-x_{\partial A}}\)\,,
\end{eqnarray}
where $w_\mu:=u_\mu/\beta$ is the vorticity vector and $\hat{n}^1, \hat{n}^2$ are two mutually orthogonal unit vectors which are perpendicular to the boundary entangling surface $\partial A$. 

A perfect fluid, like the one described by the stress tensor (\ref{bdy-SET-brane}) obeys the following equations of motion \eqref{div-T}
\bea\label{PF-eqs-body}
\partial_\mu \log\beta-du_\mu (u_\lambda \partial_\lambda)\log \beta+u_\mu (\partial_\lambda u_\lambda)+(u_\lambda \partial_\lambda )u_\mu=0\,,
\eea
which are simply a consequence of the conservation equation $\partial_\mu T^{\mu \nu}=0$. As we have discussed above, the bulk solution suggests an inverse temperature of the boundary fluid is $\beta_n(x)=n \beta(x)$, but we will usually separate the dependence on $n$ for simplicity of the analysis and refer mostly to $\beta(x)$. This is because \eqref{PF-eqs-body} are invariant under a constant rescaling of $\beta(x)$.

Equations \eqref{PF-eqs-body} can also be seen to reduce to a single non-linear differential equation (\ref{eq-fluid-scalar}) for a scalar function $\alpha$ defined through  $w_\mu:=\partial_\mu \alpha$, when one assumes the vorticity condition (\ref{circulation-source}). This is
\begin{eqnarray}\label{fluid-eq-alpha}
\partial^2 \alpha + (d-2)\, \partial_\mu \alpha\partial^\mu \log\(|\partial \alpha|\) =0\,, \quad{\rm where}\quad u_\mu =\frac{\partial_\mu \alpha}{|\partial \alpha|}\,,\quad {\rm and}\quad \beta=\frac{1}{|\partial \alpha|}\,.
\end{eqnarray}
The vorticity is zero everywhere in the Euclidean geometry except at the boundary of the entangling region where we put the source (\ref{circulation-source}) which generates a constant circulation. In terms of $\alpha$ such condition reads
\begin{eqnarray}\label{circulation-beta}
\oint_\Gamma d \alpha =\oint_\Gamma \partial_\mu \alpha \,dx^\mu=\oint_\Gamma \frac{u_\mu}{\beta} \,dx^\mu=1\,,
\end{eqnarray}
where the path $\Gamma$ links once the boundary of the entangling region with the same orientation as the one of the fluid flow. See appendix \ref{App:perfect-fluids} for a review of these results.  This implies that our ansatz depends only on the scalar function $\alpha$, constrained to satisfy \eqref{circulation-beta} and \eqref{fluid-eq-alpha}. The gradient of this function is smooth out of the entangling surface while this angular function itself must have a jump (or is defined in patches) as we circulate around  the entangling surface.  

Given the above ansatz, we would like to evaluate the refined R\'enyi entropy that follows from it. We will follow Dong's prescription \eqref{Dong-prescription}, which in this context requires to evaluate the area of the back-reacted brane, which is located at $z=z_h(x)$. This is easily done from \eqref{ansatz-2}, where the brane is sitting at $\zeta=1$, which leads to 
\bea\label{Renyi-zero-ansatz}
\tilde{S}_{n\to 0}=\frac{L^{d-1}}{4G_N}\int_\Sigma \frac{ \sqrt{|\Xi|}\, d^{d-1} \sigma}{z_h^{d-1}(x)}\,,
\eea
where $|\Xi|:=\det(\Xi_{ij})$, and $\Xi_{ij}$ is the induced metric (from the brane geometry) over a (hyper) surface $\Sigma$ homologous to the entangling region
\begin{equation}\label{Xi-metric}
\Xi_{ij}=(\delta_{\mu\nu}-u_\mu u_\nu)\frac{dx^\mu_\sigma}{d\sigma^i} \frac{dx_\sigma^\nu}{d\sigma^j}\;,
\end{equation}
obtained after stripping off the Weyl factor $L^2/z^2_h(x)$. The coordinates $\sigma^i$ parametrizes such surface. We would like to rewrite \eqref{Renyi-zero-ansatz} in terms of geometric quantities referring  to the boundary Euclidean flat geometry $\delta_{\mu \nu}$. First, the let us call $\gamma_{ij}$ the induced metric on the surface $\Sigma$, from the ambient metric $\delta_{\mu\nu}$, thus
\bea
\gamma_{ij}=\delta_{\mu\nu}\frac{dx_\sigma^\mu}{d\sigma^i} \frac{dx_\sigma^\nu}{d\sigma^j}\,, \quad{\rm similarly, \,\, we\,\,define\,\,}\gamma^{ij}{\rm \,\, such \,\,that }\quad \gamma^{ij}\gamma_{jk}=\delta^i_{\,k}\,.
\eea 
It is also convenient to define the vector $u^\sigma_i:=u_\mu \frac{dx_\sigma^{\mu}}{d\sigma^i}$, which is the projection of $u^\mu$ along the surface $\Sigma$. Thus, we get the following expression for \eqref{Xi-metric}
\bea
\Xi_{ij}=\gamma_{ij}-u^\sigma_i u^\sigma_j\,.
\eea
The determinant of the above metric, satisfies the following relations
\bea
|\Xi|=|\gamma| \, \det(\delta^{j}_{\, k}-\gamma^{jl} u^\sigma_l u^\sigma_k )\quad\to \quad \sqrt{|\Xi|}=\sqrt{|\gamma|}\,\sqrt{1-|u^\sigma |^2}\,.
\eea
The final simplification comes from the fact that the vector $u^\mu$ has norm one, and therefore, its normal component can be obtained from the norm of the tangent component $u_\sigma$, as $u^\mu \hat{n}_\mu=\pm \sqrt{1-|u^\sigma|^2}$. Therefore, \eqref{Renyi-zero-ansatz} can be seen as the flux over the surface $\Sigma$ of the conserved current $J^\mu$: 
\bea\label{Renyi-0-ansatz-2}
\tilde S_{n\to0}=\int_\Sigma \sqrt{|\gamma|}\,d^{d-1}\sigma\, \hat{n}_\mu J^\mu\,,
\eea
\bea
{\rm where }\quad J^\mu=\frac{ \sigma_s}{n^{d-1}}\, \frac{u^\mu}{\beta(x)^{d-1}}\,, \quad{\rm with }\quad \sigma_s=\frac{L^{d-1}}{4 G_N  }\(\frac{4\pi}{d}\)^{d-1}\,,
\eea
and we have used the relation $z_h(x)= nd \beta(x)/4\pi$. Importantly, conservation of $J^\mu$ is guaranteed from the fluid equations, see Appendix \ref{App:perfect-fluids}.

In summary, our ansatz correctly reproduces the features found in the known brane solution dual to the state $\tilde{\rho}_n$ associated to a spherical and/or Rindler region, and correctly generalize it to arbitrary entangling regions. In particular, it incorporates the correct conical singularity at the location of the brane. It also produces self consistent boundary relations such as \eqref{bdy-SET-brane}  and \eqref{Renyi-0-ansatz-2}. This ansatz however, is not a priori a solution to the Einstein's equations with negative cosmological constant, except when $u$ and $z_h$ are constant. We are interested in exploring under which conditions the above ansatz solves the Einstein's equations in the following subsection. We expect this to be the case at leading order in a perturbative small $n$ expansion provided the scalar function $\alpha$ satisfy the fluid equation (\ref{fluid-eq-alpha}).

\subsection{Check of the geometric ansatz in \texorpdfstring{$d>2$}{d > 2} \label{ansatz-d}}

In this section, we will check whether our brane ansatz \eqref{f3} is an actual solution to the Einstein's equations in the appropriate $n\to 0$ limit for $d>2$. For this purpose, it is convenient to work with the metric (\ref{ansatz-2}) and perform the following bulk coordinate redefinition:
\begin{eqnarray}
\rho=\(\frac{1-\sqrt{1-\zeta^d}}{1+\sqrt{1-\zeta^d}}\)^{\frac{2}{d}}\,, \quad{\rm where}\quad \rho \in\[\frac{q^2_n\, \epsilon^2}{16^{1/d}\beta^2(x)},1\]\,,\quad{\rm with}\quad q_n=\frac{4\pi}{nd}\,,
\end{eqnarray}
and the lower end in the range of the coordinate $\rho$ is obtained in the $\epsilon \to 0$ limit.\footnote{ The various terms in (\ref{ansatz-2}) take the following form
\begin{eqnarray}
\frac{d\zeta^2}{\zeta^2(1-\zeta^d)}=\frac{d\rho^2}{4\rho^2}\,, \quad \frac{1}{\zeta^2}=\frac{1}{\rho}\(\frac{1+\rho^{\frac d2}}{2}\)^{\frac{4}d}\,,\quad{\rm and} \quad \frac{(1-\zeta^d)}{\zeta^2}=\frac{1}{\rho}\(\frac{1+\rho^{\frac d2}}{2}\)^{\frac{4}d-2}\(\frac{1-\rho^{\frac d2}}{2}\)^2\,,
\end{eqnarray}
in terms of $\rho$.} Thus, our ansatz has an exact Fefferman-Graham form, this is 
\begin{eqnarray}\label{FG-metric}
ds^2=G_{ab}(\tilde{x})d\tilde{x}^ad\tilde{x}^b=L^2\frac{d\rho^2}{4 \rho^2}+\frac{L^2}{\rho}g_{\mu \nu}(x,\rho) dx^\mu dx^\nu \,,
\end{eqnarray}
where $G_{ab}(\tilde{x})$ is the $d+1$ dimensional metric, with $\tilde{x}^{d}=\rho$ and $\tilde{x}^\mu=x^\mu$ with $\mu=\{0,\cdots, d-1\}$, while $g_{\mu \nu}(x,\rho)$ is an effective $d$-dimensional metric given by\footnote{As emphasized around \eqref{circulation-beta}, our ansatz depends on the single scalar function $\alpha$. Thus, equation \eqref{g-metric}, can be further simplified to 
\begin{eqnarray}\label{anzats-metric-eqs}
g_{\mu \nu}(x,\rho)=q_n^2\(\frac{1+\rho^{\frac{d}{2}}}{2}\)^{\frac{4}d}\[  \( \partial \alpha\)^2 \delta_{\mu \nu}-\frac{4 \rho^{\frac{d}{2}} }{\(1+\rho^{\frac{d}{2}}\)^2} \partial_\mu \alpha \partial_\nu \alpha\]\,,
\end{eqnarray} 
after replacing $u^\mu$ and $\beta$ as functions of $\alpha$ via \eqref{fluid-eq-alpha}. However, as we will see, it will be more convenient to work with the more general expression \eqref{g-metric}.}
\begin{eqnarray}\label{g-metric}
g_{\mu \nu}(x,\rho)=\frac{q_n^2}{\beta^2}\(\frac{1+\rho^{\frac{d}{2}}}{2}\)^{\frac{4}d}\[ \delta_{\mu \nu}-\frac{4 \rho^{\frac{d}{2}} }{\(1+\rho^{\frac{d}{2}}\)^2} u_\mu u_\nu\]\,,
\end{eqnarray}
and $q_n\sim 1/n$, and thus, it is a large parameter in our regime of interest. From this perspective the metric $G_{ab}(\tilde{x})$ has a particular but important structure:
\bea\label{Gorders}
G_{dd}(\tilde{x})\sim {\cal O}(1),\quad  G_{d\mu}(\tilde{x})\sim {\cal O}(1), \qquad {\rm while } \qquad G_{\mu \nu}(\tilde{x})\sim {\cal O}\(\frac{1}{n^2}\)\,.
\eea

The Einstein's equations with negative cosmological constant $\Lambda=-d(d-1)/2L^2$ can be written in the form $R_{ab}+d \,G_{ab}/L^2=0$.\footnote{The Einstein's equations with cosmological constant $\Lambda$ are:
 \bea\label{EEqs-L}
 R_{ab}-\frac{R}{2} G_{ab}+\Lambda G_{ab}=0\, \qquad {\rm implies} \quad R=\frac{2(d+1)}{(d-1)}\Lambda\,,
 \eea
where the right hand side of (\ref{EEqs-L}) is obtained from its left hand side after taking its trace. If one further replace the trace back into the left hand side of (\ref{EEqs-L}), and uses the value of $\Lambda$, one arrives at the simpler equation $R_{ab}+d \,G_{ab}/L^2=0$.} In the coordinate system (\ref{FG-metric}), these  equations can be re-written in terms of the auxiliary metric $g_{\mu\nu}(x,\rho)$. Separating the above equations into boundary-boundary, boundary-bulk and bulk-bulk components, we get the following set of partial differential equations \cite{deHaro:2000vlm}: 
\begin{eqnarray}\label{EEqs-bdy}
\rho\[2g''-2g' g^{-1} g' +\tr\(g^{-1} g'\) g'\]+{\rm Ric}(g)-(d-2)g'-\tr\(g^{-1} g'\) g&=&0\,, \\ \label{EEqs-vector}
\nabla_\mu \tr\(g^{-1} g'\)-\nabla^\nu g'_{\mu \nu}&=&0 \,,\\ \label{EEqs-scalar}
 \tr\(g^{-1} g''\)- \frac12\tr\(g^{-1} g'g^{-1} g'\)&=&0\,,
\end{eqnarray}
where Ric$(g)$ is the Ricci tensor for the metric $g$, $\nabla_\mu$ denotes its associated covariant derivative and $'$ denotes the derivative with respect to $\rho$.

 We want to analyze these equations perturbatively in the small $n$ regime.
For that purpose, it is important to keep track of the order of the metric components $G_{ab}$ as a function of $n^2$, (\ref{Gorders}). Let us take (\ref{EEqs-bdy}) which is equivalent to $R_{\mu \nu}+d \,G_{\mu\nu}/L^2=0$. Since in our ansatz, $G_{\mu \nu}\sim {\cal O}(1/n^2)$, at leading order in $n\to0$, we expect  (\ref{EEqs-bdy}) to be valid only at order ${\cal O}(1/n^2)$, and any ${\cal O}(1)$ contributions could only be accounted for by modifying our ansatz, such that the components  $G_{\mu \nu}$ include ${\cal O}(1)$ corrections. Given that  $g\sim {\cal O}(1/n^2)$, $g^{-1}\sim {\cal O}(n^2)$, we deduce that at leading order in the $n\to 0$ limit the only term in equation (\ref{EEqs-bdy}) that is not of order ${\cal O}(1/n^2)$  is Ric($g$). Therefore, the Einstein's equations (\ref{EEqs-bdy}) in the $n\to 0$ limit reduce to 
\bea\label{EEqs-bdy-2}
\rho\[2g''-2g' g^{-1} g' +\tr\(g^{-1} g'\) g'\]-(d-2)g'-\tr\(g^{-1} g'\) g=0\,. 
\eea
On the other hand, equations (\ref{EEqs-vector}) follow from $R_{\mu \rho}+d \,G_{\mu\rho}/L^2=0$. Since our ansatz has $G_{\mu\rho}=0$, we expect corrections to these components to appear only at order ${\cal O}(n^2)$. There are no terms of that order in equation (\ref{EEqs-vector}), thus, we conclude that those equations do not simplify at leading order in the $n\to 0$ limit. Finally, equation (\ref{EEqs-scalar}) follows from $R_{\rho \rho}+d \,G_{\rho\rho}/L^2=0$, and since $G_{\mu\rho}\sim {\cal O}(1)$ we expect their corrections to appear at order $ {\cal O}(n^2)$. Once again this implies that equation (\ref{EEqs-scalar}) does not simplify.  In summary, if our ansatz is to be a good approximation to the brane geometry in the $n\to 0$ limit, it would need to solve the approximated Einstein's equations (\ref{EEqs-vector}), (\ref{EEqs-scalar}) and (\ref{EEqs-bdy-2}).

 Now, we would like to evaluate our anzats in the Einstein's equations (\ref{EEqs-vector}), (\ref{EEqs-scalar}) and (\ref{EEqs-bdy-2}). We can compute the determinant and inverse of $g$, which have the form
\begin{eqnarray}\label{det-inverse}
 \det(g)=\frac{1}{16}\frac{q_n^{2d}}{\beta^{2d}} \(1-\rho^d\)^2\,, \qquad g^{\mu \nu}(x,\rho)=\frac{\beta^2}{q_n^2 } \(\frac{1+\rho^{\frac{d}{2}}}{2}\)^{-\frac{4}d}\[ \delta_{\mu \nu}+\frac{4 \rho^{\frac{d}{2}} \,  u_\mu u_\nu}{\(1-\rho^{\frac{d}{2}}\)^2}\]\,. 
\end{eqnarray}
These expressions facilitate the evaluation of the various terms in the Einstein's equations. Using these explicit expressions, and the relation $\tr \(A^{-1}d A\)=d\log\(\det A\)$ we find 
\begin{eqnarray}
 \tr\(g^{-1} g'\)=- \frac{2d\rho^{d-1}}{1-\rho^d}\,, \,\,\,\, {\rm and}\,\,\,\,\tr\(g^{-1} g' g^{-1} g'\)=2\, \tr\(g^{-1} g''\)=- \frac{4d\rho^{d-2}(1-d-\rho^{d})}{\(1-\rho^d\)^2}\,, 
\end{eqnarray}
which implies that equation (\ref{EEqs-scalar}) is satisfied identically by our anzats independently of $u_\mu$, $\beta$. Similarly, it is easy to show that our anzats satisfy (\ref{EEqs-bdy-2}) exactly. The remaining equations (\ref{EEqs-vector}) can be simplified. First, the term $\nabla_\mu \tr(g^{-1}g')=\nabla_\mu\partial_\rho \log\det(g)$ vanishes identically, since the determinant of $g$ factorizes into a function of $\rho$ times a function of $x$ (\ref{det-inverse}). The other term can be evaluated, and give rise to a differential equation for $\beta$ and $u_\mu$, this is 
\begin{eqnarray}
\nabla^\nu g'_{\mu \nu}=-\frac{2d \rho^{\frac d2-1}}{1-\rho^d}\(\partial_\mu \log\beta-du_\mu (u_\lambda \partial_\lambda)\log \beta+u_\mu (\partial_\lambda u_\lambda)+(u_\lambda \partial_\lambda )u_\mu\,\)=0.
\end{eqnarray}
The vanishing of the above equation is guaranteed provided $\beta$ and $u_\mu$ satisfy the equations of a perfect fluid \eqref{PF-eqs-body}. In fact, since $u_\mu$ and $\beta$ are related through $\partial_\mu\alpha$, the resulting equations in terms of $\alpha$ are simply
\begin{eqnarray}\label{Einstein-to-fluid}
\nabla^\nu g'_{\mu \nu}=-\frac{2d \rho^{\frac d2-1}}{1-\rho^d}\frac{\partial_\mu \alpha}{\(\partial \alpha\)^2}\(\partial^2 \alpha - \frac{(d-2)}{2}\,\(\partial_\lambda \alpha\)  \partial_\lambda \log(\partial \alpha)^2\)=0\,, 
\end{eqnarray}
which agree with our simplified fluid equation (\ref{fluid-eq-alpha}).

In summary, plugging our anzats into the Einstein's equations reduces to a set of partial differential equations which are satisfied provided the boundary functions $u^\mu$, $\beta$ satisfy the equations of a perfect fluid.

\subsection{The geometric ansatz for \texorpdfstring{$d=2$}{d = 2}\label{Check2d}}

Two boundary dimensions is very special, namely due to the absence of gravitons in three dimensions.
Then, it is interesting to explore whether our ansatz becomes an exact solution for $n\rightarrow 0$ in the two dimensional case. We show that whilst solving locally the equations of motion exactly for all $n$, the presence of discontinuities in the fluid solutions induce topological obstructions in being a true solution to the equations \eqref{Grav+Brane-Action}. We explore these properties of our solution using the induced boundary $T_{\mu\nu}$ showing that it has an inadequate pole structure induced by the zeroes of $\beta$. We also show that these topological defects are subleading in $n$ with respect to Dong's brane and vanish as $n\to0$ which allows our solution to be taken again as a good ansatz to compute $\tilde S_{n}$ in this limit.

In $d=2$ the ansatz (\ref{FG-metric}), takes the explicit form
\begin{align}
\label{ansatz-2d}
ds^2=& L^2\left(\frac{d\rho^2}{4\rho^2}+\frac{q_n^2}{\rho}\left[\(\frac{1+\rho}{2}\)^2(\partial_y\alpha)^2 +\(\frac{1-\rho}{2}\)^2(\partial_x\alpha)^2 \right] dx^2\nonumber\right.\\
& \left.-q_n^2 (\partial_x\alpha) (\partial_y\alpha) dx dy+\frac{q_n^2}{\rho}\left[\(\frac{1+\rho}{2}\)^2(\partial_x\alpha)^2 +\(\frac{1-\rho}{2}\)^2(\partial_y\alpha)^2 \right] dy^2 \right)\,.
\end{align}
As shown in section \ref{ansatz-d}, at leading order in the $n\to 0$ limit, the above ansatz satisfies the Einstein's equations whenever $\alpha$ obeys the equations of a perfect fluid (\ref{fluid-eq-alpha}) with circulation condition \eqref{circulation-beta}. However, the difference between the full Einstein's equations  (\ref{EEqs-scalar}), (\ref{EEqs-vector}), (\ref{EEqs-bdy}) and the leading order ones is the term Ric($g$), c.f. (\ref{EEqs-bdy-2}). In other words, if it happens that Ric($g$) in the above case vanishes, we would find that the full non-linear Einstein's equations are locally solved by our ansatz. Evaluating Ric($g$) for the effective two dimensional metric $g(x,\rho)$ that can be read from (\ref{ansatz-2d}), one gets:
\bea
{\rm Ric}(g)_{\mu \nu}=F(x,\rho)g_{\mu \nu}\,, 
\eea
where 
\begin{align*}
F(x,\rho)=& \frac{4\rho}{q_n^2(1-\rho)^2\(\partial \alpha\)^6} \Big( \[(\partial_x\alpha )^2-(\partial_y\alpha )^2\]\[(\partial^2_x\alpha )^2-(\partial^2_y\alpha )^2\] +(\partial_x\alpha)(\partial_y\alpha) (\partial_x\partial_y \alpha)(\partial^2 \alpha) \\
&-(\partial_x\alpha)^3\partial_x(\partial^2 \alpha)-(\partial_x\alpha)(\partial_y\alpha)\[(\partial_x\alpha)\partial_y(\partial^2 \alpha)+(\partial_y\alpha)\partial_x(\partial^2 \alpha)\]-(\partial_y\alpha)^3\partial_y(\partial^2 \alpha) \Big)\,.
\end{align*}
The above quantity vanishes, and therefore, Einstein's equations hold exactly provided the scalar function $\alpha$ satisfies the Laplace equation 
\begin{eqnarray}\label{fluid-eq-alpha-d=2}
\partial^2\alpha=\partial_x^2\alpha+\partial_y^2\alpha=0\,,
\end{eqnarray}
which is precisely the fluid equation for the boundary $d=2$ fluid (\ref{fluid-eq-alpha}). The associated boundary solution to the fluid equations for $N$ disjoint intervals with boundaries $\{l_i,r_i\}$ and with the appropriate vorticity conditions \eqref{circulation-beta} is known to be given by
\begin{equation}\label{Sol-Fluid-d=2}
\alpha(z)={\frac{i}{2\pi}}\log\(\prod_{i=1}^N \frac{z-l_i}{r_i-z}\)\quad\Rightarrow\quad\beta(x)=\frac{1}{|\partial_z \alpha(z)|}\,. 
 \end{equation}
where we use complexified coordinates $z=x+iy$. Thus, using the formula \eqref{Renyi-0-ansatz-2} and evaluating the integral on the $y=0$ surface, we obtained for the refined R\'enyi entropy 
\begin{eqnarray}\label{Sn-gen-2d-2}
\tilde{S}_n=\frac{c}{3\,n}\!\!\( \sum^N_{i,j} \log|r_j-l_i|-\sum^N_{i<j} \log|r_j-r_i|-\sum^N_{i<j} \log|l_j-l_i|-N\log\epsilon \),\quad c=\frac{3 L}{2G_N}\,,
\end{eqnarray}
where we have regulated the entropy calculation in the usual way, and used the Brown-Henneaux relation to express the R\'enyi entropies in terms of the central charge.

However, in order to claim to find an exact solution for our problem of interest \eqref{Grav+Brane-Action}, we would need to check that no other defects have appeared in the bulk. We will see through the boundary stress energy tensor that this is not the case for arbitrary $n$ but that it holds for $n\to0$. The boundary stress tensor is easy to compute from the metric (\ref{f3}) by the usual techniques \cite{Fefferman:95, Balasubramanian:1999re, deHaro:2000vlm}.
Explicit evaluation leads to the traceless and conserved form
\begin{eqnarray}\label{stress-T-ansatz-2d}
T_{\mu \nu}=\frac{L\pi }{4G_N}\(\frac{1}{n^2} t_{\mu \nu} +f_{\mu \nu}\)\,,\qquad\qquad ds^2=dx_\mu dx^\mu\;,\qquad\qquad (d=2)\;, 
\end{eqnarray}
where we have used the relation $q_n=2\pi/n$ and repeated the boundary metric for future convenience. The tensor $t_{\mu\nu}$ and its subleading ${\cal }O(n^0)$ correction $f_{\mu \nu}$ are independently traceless, conserved and have the form of a perfect fluid:
\bea
t_{\mu \nu}=\frac{\(\delta_{\mu \nu}-2u_\mu u_\nu\)}{\, \beta^2}\,, \qquad {\rm and}\qquad f_{\mu \nu}=\frac{\(\delta_{\mu \nu}-2\tilde{u}_\mu\tilde{u}_\nu\)}{\tilde{\beta}^2}\,.
\eea
The fluid variables $u^\mu$, $\beta$ which satisfies the fluid equations \eqref{PF-eqs-body} together with the circulation condition \eqref{circulation-beta} can be obtained from \eqref{Sol-Fluid-d=2}. On the other hand, the fluid variables $\tilde{u}^\mu$, $\tilde{\beta}$ are given by 
 \bea
 \tilde{\beta}=\frac{\beta}{|\partial \beta|}\,, \quad{\rm and}\quad \tilde{u}_\mu 
=\pm \epsilon_{\mu \nu} \frac{\partial_\nu \beta}{|\partial \beta|}\,,
 \eea
and likewise satisfy the fluid equations. The tensors $t_{\mu\nu}$ and $f_{\mu\nu}$ have some analytic properties that are worth mentioning. For that purpose, it is convenient to express them in holomorphic coordinates $\{z,\bar z\}$, where  $t_{zz}$ and $f_{zz}$ are expressed as 
\begin{equation}
t_{zz}=-\frac{1}{2}(\partial_z\al)^2=-\frac{1}{8\pi ^2}\left[\sum_{i=1}^N \(\frac{1}{r_i-z}+ \frac{1}{z-l_i} \)\right]^2\,,
\end{equation}
 \begin{equation}
f_{zz}=-\frac{1}{2}\(\frac{\partial^2_z\al}{\partial_z \al}\)^2\,,\quad{\rm where}\quad \frac{\partial^2_z\al}{\partial_z \al}=-\frac{\sum_{i=1}^N \(\frac{1}{(z-l_i)^2}-\frac{1}{(z-r_i)^2} \)}{\sum_{i=1}^N \(\frac{1}{z-l_i} -\frac{1}{z-r_i}\)}\,,
\end{equation}
and the analogous for the anti-holomorphic parts. As shown in the above expressions,  $t_{zz}$ has simple and double poles at the vortexes (in the replica framework, those correspond to the twist operator insertions) while the $f_{zz}$ has additional poles. More specifically, one can show that $f_{zz}$ retains the double and single poles at the vortices, but it develops $2N-2$ new poles in the complex plane away from the real axis. These points appear whenever $|\partial \alpha|=0$ or equivalently $\beta\to\infty$, and are induced by the circulation condition on the fluid velocities for non-trivial entangling topological regions, which necessarily imply the existence of discontinuities in the fluid velocities. 
These extra poles in the $T_{\mu\nu}$ argue against our $d=2$ ansatz \eqref{ansatz-2d} being an exact solution to the set of equations \eqref{Grav+Brane-Action}. 
On the other hand, global conformal invariance implies that $T_{zz}\sim 1/z^4$ in the $z\to \infty$ limit. Such behavior is obeyed by $t_{zz}$ but it is not the case for $f_{zz}$ which also disregard our geometry as the actual dual solution to our boundary problem.

Another way to highlight the problems induced by taking our ansatz as an exact saddle in $d=2$ for the complete boundary is to choose a different boundary metric of the same conformal family in which the stress tensor looks simpler. One can show that using a flat $\zeta=\epsilon$ cutoff on \eqref{ansatz-2} one lands at
\begin{eqnarray}\label{T-ansatz-non-flat}
T_{\mu \nu}=\frac{L\pi }{4G_N}\frac{\(\delta_{\mu \nu}-2u_\mu u_\nu\)}{ n^2\, \beta^2}\,,\qquad\qquad ds^2=\frac{dx_\mu dx^\mu}{\beta^2}\;,\qquad\qquad (d=2)\;, 
\end{eqnarray}
which should be contrasted with \eqref{stress-T-ansatz-2d}. We can see that the stress tensor does indeed take the form of a perfect fluid with poles only induced by the vertices as long as the metric does not present singularities, which it does whenever $\beta$ is singular. We already know that $\beta=0$ at the vertices by construction, but one can also check that the fluid equations induce points of divergent $\beta\to\infty$ temperature whenever the gradient of $\alpha$ is ill defined as explained in the previous paragraph. The extra piece $f$ in \eqref{stress-T-ansatz-2d} can be seen to appear from the conformal map that removes the factor $\beta^2$ from the denominator above involving gradients of $\ln\beta$, see \cite{Skenderis:2000in}. This restricts the validity of our solution within each of these domains where the local fluid variables $\{\beta,u_\mu\}$ are single-valued.  

We also mention that the subtraction point for the zero energy has already been fixed to be pure AdS, appearing as constant $z_h\to\infty$. As such, despite $f_{zz}$ being an $O(n^0)$ contribution to the stress tensor, it should not be removed by moving the subtraction point of zero energy. In fact, explicit computation shows that for the Rindler solution, both $t$ and $f$ are of the same functional form and together they recover the full stress tensor of the exact solution eq. \eqref{Conf-Transf}.

In any case, these deviations from the expected boundary stress tensor $t_{\mu\nu}$ are subleading as $n\to0$. We thus conclude that in $d=2$, where our ansatz exactly meet the equations \eqref{Grav+Brane-Action} locally, it fails to meet the required boundary conditions on the full boundary manifold, except at the $n\to0$ limit. This is, our ansatz is on the same footing in any dimensions $d\geq2$ as a leading order ansatz in the $n\to0$ limit, and \eqref{Sn-gen-2d-2} must also be taken in that sense.\footnote{Notice that coming from the ansatz \eqref{f3}, which is a leading order solution in an $n\to0$ limit, we no longer get $T_{\mu \nu}=0$ for $n=1$, c.f. \eqref{Conf-Transf}. The reason for this is that the physical reference point is always the pure AdS solution which in this context is constant $z_h\to\infty$ for which we do get $T_{\mu \nu}=0$, rather than $n\to1$. The exact solution must recover $T_{\mu \nu}=0$ for $n=1$.}

\section{Holographic R\'enyi entropies in two dimensions \label{Sec:d=2}}

In this section, we first discuss the problem of the R\'enyi entropies in $d=2$ for general CFT's. We argue the limit $n\rightarrow 0$ should be universal and correspond to a fluid that is locally thermal at the so-called entanglement temperature. These temperatures determine the local terms in the modular Hamiltonian.  
Then, following Ba\~nados \cite{Banados:1998gg}, we construct the dual geometries to the state $\rho^n/\tr{\rho^n}$ for holographic models from the knowledge of the expectation value of the boundary stress tensor in the aforementioned state. We reproduce the well known holographic R\'enyi entropy of two dimensional Rindler for arbitrary $n$ eq. \eqref{Sn-rindler} in Sec. \ref{2dRindler}, and derive $\tilde S_{n\to0}$ for an arbitrary number of intervals eq. \eqref{Sn-gen-2d} in Sec. \ref{2dNIntervals}. 

\subsection{Entanglement temperatures}

Let a general multi-interval region in $d=2$ (at the line $t=0$) be $A=\cup_{i=1}^N (l_i,r_i)$. In a CFT the modular Hamiltonian $K$ of the vacuum is non local except for the case of a single interval. However, it was conjectured in \cite{Arias:2016nip,arias2017anisotropic} that for multi-intervals there is a local term in $K$ having the same universal structure for any CFT. This writes 
\be
K_{\rm loc}=\int dx \, \beta(x)\, T_{00}(x)\,,\label{loc}
\ee
where the ``local inverse temperatures'' are
 \be
\beta(x)=\frac{2\pi}{\sum_{i=1}^n \left(\frac{1}{x-l_i}-\frac{1}{x-r_i}\right)}\,. \label{beta00}
\ee  
The modular Hamiltonian for multi-interval regions in $d=2$ is known explicitly for free massless scalar and fermions \cite{casini2009reduced,arias2018entropy}, and the local term agrees with (\ref{loc}).   

A clear operational definition of these entanglement temperatures is that they measure the ratio between the expectation value of the modular Hamiltonian and the energy for very localized high energy excitations:
\bea
\beta(x)=\lim_{E\rightarrow \infty\,, \Delta x\rightarrow 0}\frac{\langle \psi| K |\psi\rangle}{\langle \psi| H |\psi\rangle}\,.
\eea
Here $|\psi\rangle$ is an excitation above the vacuum localized in $\Delta x$ around the point $x$ and has energy $E$. This limit can be argued to exist based on monotonicity of the relative entropy \cite{Arias:2016nip}.

We will now show the universality of these entanglement temperatures in a more formal way. To test the entanglement temperatures we will be using a chiral field operator acting on the vacuum as the excited state probe.   
We would like to compute (for a real chiral field $\phi$)
\bea
\beta(x)=\lim_{y\to x}\frac{\langle 0|\phi(x) K \phi(y)|0\rangle}{\langle 0|\phi(x) H \phi(y)|0\rangle}\,.
\eea
This measures the ratio between the modular energy $K$ and the ordinary energy $H$. As a regularization we set $x\neq y$ and  take the limit $x\rightarrow y$.\footnote{The structure of the singularity turns out to be the same in numerator and denominator, so we can equivalently take fields smeared in a short interval around a point.}  
The modular operator $K$ can be obtained from the modular flow $\Delta^{i\, \tau}=e^{-i K \tau}$ with $K=-\log\Delta$.  We get the entanglement temperature as a limit of a modular evolved correlator  
\bea\label{beta-chiral-2}
\beta(x)=\lim_{y\to x} \lim_{\tau\to 0}-\frac{ \partial_\tau \langle 0|\phi (x) \Delta^{i \tau} \phi(y)|0\rangle}{\partial_{y^0}\langle 0|\phi(x) \phi(y)|0\rangle}\,.\label{limit}
\eea

For a general chiral field $\phi$ in a  CFT, S. Hollands obtained a general structure of the modular evolved correlator \cite{Hollands:2019hje}. From a clever use of the KMS condition for the modular flow and analyticity in the Euclidean plane Hollands maps the problem into a Riemann-Hilbert problem in the plane with a cut at $A$.  His expression will be enough to get the entanglement temperatures by the limit (\ref{limit}). 

Take a chiral field $\phi$ of dimension $h$. By locality $h$ is a half integer number. The field is normalized such that 
\be
\langle 0|\phi(x)\phi(y)|0\rangle= \frac{e^{-i\pi\, h}}{(x-y-i 0^+)^{-2 h}}\,.
\ee
  Defining 
\be
\Pi_l(x)=\prod_{j=1}^N (x-l_j)\,,\qquad \Pi_r(x)\equiv \prod_{i=1}^N (x-r_i)\,,\qquad Z(x)=\frac{1}{2\pi}\ln(-\Pi_l(x)/\Pi_r(x))\,,
\ee
the result of \cite{Hollands:2019hje} is that 
\bea \label{modular-chiral}
\langle 0|\phi(x)\Delta^{i\,\tau} \phi(y)|0\rangle =\left(\Pi_l(x)\Pi_r(x)\Pi_l(y)\Pi_r(y)\right)^{-h} 
 \, \sum_{i,j=0}^{2(N-1)h} \,\, \hat{c}_{ij}\left[\tau+Z(x)-Z(y)\right]\, x^i\, y^j\,.
\eea
This is a distribution in $(x,y)\in A\times A$ 
where  $\hat{c}_{ij}(\tau)$ is analytic in the strip $\{ t\in \mathbb{C}|-1<\Im\, t <0\}$. There, it satisfies a bound $
|\hat{c}_{ij}(\tau)| \leq \left[\sin(\pi \Im\, \tau )\right]^{-2h}$. 
In the limit $\tau\rightarrow -i 0^+$ we have, matching the distribution on both sides of (\ref{modular-chiral}),
\be\label{2pf-chiral}
\sum_{i,j=0}^{2(N-1)h} \hat{c}_{ij}[Z(x)-Z(y)-i0^+]\, x^i \,y^j=(-1)^h\,\left( \frac{Q(x,y)}{2\,\sinh \pi(Z(x)-Z(y)-i0^+)}\right)^{2h}\,,
 \ee
 where
\bea
\label{Q}
Q(x,y)=\frac{\prod_{j=1}^N (x-l_j) (y-r_j)-\prod_{j=1}^N (y-l_j)(x-r_j)}{x-y}\,,
\eea
is smooth as $x\rightarrow y$.

Notice that (\ref{modular-chiral}) for $\tau=-i0^+$ is the ordinary correlator, and diverges in the $x\to y$ limit, as it must be the case. This is explicit in (\ref{2pf-chiral}), from which one deduce that at least some of the $\hat{c}_{ij}[Z(x)-Z(y)]$ must be divergent in the $y\to x$ limit. On the other hand, the two point function, giving the leading divergence in the denominator of (\ref{beta-chiral-2}) can only come from the derivative of these functions. In other words, since all the other factors in (\ref{modular-chiral}) are smooth when $y\to x$ their derivative  can only give rise to a subleading contribution. This means that the leading divergence in the denominator of (\ref{modular-chiral}) is given by 
\bea
\partial_{y}\langle 0|\phi(x) \phi(y)|0\rangle \approx  -(\Pi_l(x)\Pi_r(x))^{-2h} \sum_{i,j=0}^{2(N-1)h} \hat{c}'_{ij}[Z(x)-Z(y)-i0 ] x^{i+j} \partial_x Z(x)\,, 
\eea
where  we have $\partial_{y^0}=\partial_y$ for the chiral field. On the other hand the numerator approaches 
\bea
 \lim_{\tau\to 0} \partial_\tau \langle 0|\phi (x) \Delta^{i \tau} \phi(y)|0\rangle \approx  (\Pi_l(x)\Pi_r(x))^{-2h} \sum_{i,j=0}^{2(N-1)h} \hat{c}'_{ij} [Z(x)-Z(y)-i0] x^{i+j}\,.
\eea
Thus we see that the leading divergence of both numerator and denominator matches exactly, which leads to a closed formula for the local temperature that recover \eqref{beta00},
\bea
\beta(x)=(\partial_x Z(x))^{-1}=\frac{2\pi }{\sum_{j=1}^N (\frac{1}{x-l_j}-\frac1{x-r_j})}\,.
\eea

\subsection{Stress tensor in the limit \texorpdfstring{$n\rightarrow 0$}{n->0} \label{stress-tensor-1}}

In the replica trick computation of $\rho^n$ for integer $n$ we have $n$  sheets connected by a symmetry $Z_n$. The expectation value of the stress tensor in any of these sheets is the same and only depends on $n$ and the region $A$. This, then can be analytically continued in $n$. The holomorphic components of the stress tensor are analytic functions in the plane, except at the end points of the intervals. There, we can have at most single and double poles
\be   
T_{zz}(z)\sim -\frac{k^2_n\sigma_\varepsilon}{2\pi^2} \(\,\frac{1}{(z-r_i)^2}+\frac{p_i}{z-r_i}\)\,,\quad \quad z\sim r_i\,,\label{tete}
\ee 
and similarly around $z\sim l_i$. The coefficient in front of the double pole is 
\begin{eqnarray}
k^2_n=\frac{1-n^2}{4n^2}\,,
 \end{eqnarray}
 as can be calibrated by the case of Rindler space. 
 We defined $k_n^2$ to be positive, since we are interested in studying the regime $0<n<1$.
 As discussed in detail in \cite{Faulkner:2013yia, Headrick:2010zt} the single pole coefficients $p_i$, which are functions of $n$ and the region $A$, are theory dependent and contain the detailed information of the R\'enyi entropies. 
 These are constrained by global conformal symmetry to satisfy 
 \bea\label{p-const}
\sum_{i}p_i=0, \qquad \sum_{i}p_iz_i+2N=0,\quad {\rm and}\quad
\sum_{i} p_i z_i^2+2 \sum_{i}z_i=0\,,
\eea 
where $i$ runs over the end points of the intervals which are generically represented by $z_i$. These constrains guarantees that $T_{zz}(z)\sim 1/z^4$  when $z\to \infty$. 

Now, the $n\rightarrow 0$ limit is one of large modular temperatures. It is reasonable to expect that the state $\rho^n/\tr \rho^n$ becomes highly excited in this case, and that long distance correlations are screened and get suppressed with respect to local ones. In such a case the state is well approximated by a local thermal state with respect to the local part of the modular Hamiltonian. This means we have a local thermal-like state with inverse temperature $\beta_n=n \beta$, where $\beta$ is given by (\ref{beta00}). Using the relation for the energy density of a thermal state $\varepsilon= \sigma_\varepsilon T^d$, such expectation value (for the holomorphic component) is given by 
\begin{eqnarray}\label{2d-stress-tensor}
T_{zz}(z)=-\frac{\sigma_\varepsilon}{8 n^2\,\pi^2}\left[\sum_{i=1}^N \(\frac{1}{r_i-z}+ \frac{1}{z-l_i} \)\right]^2 \,, \qquad T_{\bar{z}\bar{z} }(\bar{z})=\overline{T_{zz}(z)}\,, \quad T_{z\bar{z}}(z,\bar{z})=0\,.
\end{eqnarray}
 Note that the double poles in this expression match (\ref{tete}) in the $n\to 0$ limit. This expression also fixes the single pole coefficients $p_i$ to be\footnote{See Appendix \ref{Faulkners-method} for a short derivation}
 \bea\label{pi-fluid}
p_i=\sum_{j\neq i}^{2N}\frac{ 2 (-)^{i+j}}{z_i-z_j} \,.
\eea
Here $z_i$ represents both the left and right endpoints of the intervals, namely $z_{2k}=r_k$, $z_{2k+1}=l_k$, with $k\in \{1, \cdots, N\}$. Importantly, the $p_i$'s in (\ref{pi-fluid}) satisfy the conditions (\ref{p-const}). This can be alternatively seen from the fact that the temperature \eqref{beta00} decays as $1/z^2$ at large $z$.  Interestingly, one can arrive at (\ref{2d-stress-tensor}) and (\ref{pi-fluid}) by following the prescription in \cite{Faulkner:2013yia}, and as a consequence to our main result \eqref{Sn-gen-2d} in the $n\to0$ limit. See Appendix \ref{Faulkners-method} for a full account of that derivation.

In two dimensions, any symmetric and traceless tensor, can be put, locally, in the form of a fluid. A simple counting of the number of degrees of freedom support this claim. Namely, a traceless symmetric two by two matrix is determined by two numbers, while a fluid in two dimensions is determined by a two dimensional unit vector and a temperature which in total require only two numbers to specify.  To identify the fluid variables it is convenient to introduce the holomorphic function
 \begin{eqnarray}\label{alpha}
\alpha(z):=a(x,y)+i\, b(x,y)\,, 
\end{eqnarray}
 where $z=x+iy$, and $(x,y)\in \mathbb{R}^2$, and write 
\be
 T_{zz}(z)=\kappa \(\frac{\partial \alpha(z)}{\partial z}\)^2 \,.
 \ee
The real scalar functions $a(x,y)$ and $b(x,y)$ satisfy the Laplace equation in $\mathbb{R}^2$, their gradients $\partial_\mu a$ and $\partial_\mu b$ are mutually orthogonal and have equal norms. The above facts follows from the holomorphicity of $\alpha(z)$. Additionally, we want the vector field $\partial_\mu a$ to flow across the individual intervals, such that it can be used to define a unit vector characterizing the velocity of a Euclidean fluid in two dimensions via 
 \begin{eqnarray}\label{fluid-rel-2d}
 u_\mu=\frac{\partial_\mu a}{|\partial a|}\,, \quad {\rm and}\quad \beta=\frac{1}{|\partial a|}\,,
 \end{eqnarray}
 where $\beta$ represents the inverse local temperature of the fluid. These identifications allows a rewritting of the stress-energy tensor as a perfect fluid:
 \begin{eqnarray}\label{stress-T-2d}
 T_{\mu \nu}=\frac{ 8 \pi^2 \kappa }{\beta^2}\(\delta_{\mu \nu}-2 u_\mu u_\nu\)\,.
 \end{eqnarray}

 However, this identification is in general only locally defined, and a holomorphic $T(z)$ may correspond to a non single valued velocity field. This happens for generic values of the single poles in (\ref{tete}). The solution for a fluid with fixed vortex strengths  at the end points of the intervals is in fact unique, and is generated by the function
\be 
\alpha(z)=\frac{i}{2\pi} \log\(\prod_{i=1}^n \frac{z-l_i}{r_i-z}\)\,.
\ee
This fixes the relation between the single and double poles to be precisely the one in (\ref{2d-stress-tensor}). Therefore, only in the $n\rightarrow 0$ limit we expect to have a perfect fluid description of the stress tensor in the full plane. Setting $ T_{zz}(z)=2k^2_n\sigma_\varepsilon \(\frac{\partial \alpha(z)}{\partial z}\)^2$ we get the result
\be
 T_{\mu \nu}=\quad \frac{ 4 k^2_n\, \sigma_\varepsilon }{\beta^2}\(\delta_{\mu \nu}-2 u_\mu u_\nu\)=\frac{  L \pi \, k^2_n}{G_N}\frac{1}{\beta^2}\(\delta_{\mu \nu}-2 u_\mu u_\nu\)\,,
 \ee
where in the last equality we used the value of $\sigma_\varepsilon$ for $d=2$, this is $\sigma_\varepsilon=L\pi/4G_N$. In this expression $u$ and $\beta$ are given by (\ref{fluid-rel-2d}). 

\subsection{Holographic \texorpdfstring{$n\rightarrow 0$}{n->0} solution in two dimensions \label{section-2d-exact}}

 The general solution to the vacuum Einstein's equations with negative cosmological constant and AdS asymptotics in three dimensions, was found by Ba\~nados \cite{Banados:1998gg} for arbitrary values of the boundary stress-energy tensor:
\be\label{eq:Banados}
ds^2 = L^2\frac{d\zeta^2}{\zeta^2}+ \frac{L^2}{\zeta^2} \left(dz+\zeta^2 \bar {\cal L}(\bar z) d \bar z \right)\left(d\bar z+\zeta^2  {\cal L}( z) dz \right)\,,
\ee
where $\zeta \in [0,\infty)$, and 
\begin{eqnarray}
{\cal L}(z) =- \frac{8\pi G_N\,T_{zz}(z)}{L}\;,\qquad \bar {\cal L}(\bar z) =- \frac{8\pi G_N\, T_{\bar z\bar z}(\bar z)}{L} \,.
\end{eqnarray}
This metric has an exact Fefferman-Graham form. We will study the above metric for the boundary stress tensor \eqref{2d-stress-tensor} after the replacing $1/n^2\to 4k_n^2$ for finite $n$ with $0<n<1$. As mentioned before, such stress tensor corresponds to the expectation value of $T_{zz}$ in the state $\rho^n/\tr \rho^n $ only in the $n\to 0$ limit. Nevertheless, we will study the geometry in the more general finite $n$ context when possible. To understand the metric better, it is convenient to use the conformal symmetry of the boundary and perform a series of boundary conformal transformations via some coordinate transformation in the metric. 

First, it is convenient to use the fact that $u_\mu/\beta$ is a gradient, which follows from (\ref{fluid-rel-2d}), to define an angular coordinate within a patch containing the part of the fluid that crosses a single interval. Notice that around each interval there is a patch that contains all the integral curves of $u_\mu$ that crosses that interval and no other integral curves. One can cover the full complex plane by joining all such patches. A natural map that does this is the following: 
\begin{eqnarray}\label{uni-map}
w(z)=\prod_{i=1}^N \(\frac{z-l_i}{r_i-z}\), \qquad{\rm where}\qquad w(z)= e^{-2\pi i\alpha (z)}\,.
\end{eqnarray} 
This map has the feature of mapping the full complex plane $z$ into $N$ copies of the complex $w$ plane. Indeed, a single $w$ plane will define one of our patches of interest. To see this, let us use polar coordinates for $w$, the relationship with $\alpha$, (\ref{uni-map}) and (\ref{alpha}):
\begin{eqnarray}\label{w-polar}
w(z)=r e^{i\theta}\quad \to \quad r=e^{2\pi b(x,y)}\,,\quad {\rm and}\quad \theta=-2\pi\, a(x,y)\,,
\end{eqnarray}
and the function $a(x,y)$ has the property that grows smoothly from $0$ to $1$ as one circles a given interval from $y=0^-$ to $y=0^+$, lower and upper part of the interval. Thus, fixing the $r$ coordinate and moving along $\theta$, is equivalent to moving along the gradient of $a$ which is the trajectory of the fluid velocity $u$. 

Performing the above coordinate transformation in (\ref{eq:Banados}) results in
\begin{eqnarray}
&&\tilde{\cal L}(w)=\frac{k_n^2}{w^2}\quad {\rm where}\quad \tilde{\cal L}(w)=\(\frac{\partial w}{\partial z}\)^{-2} {\cal L}(z(w))\,. \\
{\rm Using}\!\!\! &&dzd\bar{z}= \left|\frac{\partial w}{\partial z}\right|^{-2}dw d\bar{w}\,, \quad {\rm and}\quad {\cal L}(z)dz^2+\bar{\cal L}(\bar{z})d\bar{z}^2=\tilde{\cal L}(w)dw^2+\bar{\tilde{{\cal L}}}(\bar{w})d\bar{w}^2, \quad 
\end{eqnarray}
one can re-write (\ref{eq:Banados}) as 
\begin{equation}
ds^2=L^2\frac{d\zeta^2}{\zeta^2}+\frac{L^2}{\zeta^2}\left|\frac{\partial w}{\partial z}\right|^{-2}\!\!dw d\bar{w}+L^2\(\tilde{{\cal L}}(w)dw^2+\bar{\tilde{{\cal L}}}(\bar{w})d\bar{w}^2\)+\,L^2\zeta^2\left|\frac{\partial w}{\partial z}\right|^{2}\!\! \tilde{{\cal L}}(w)\bar{\tilde{{\cal L}}}(\bar{w}) dw d\bar{w}\,. \nonumber
\end{equation}
Using the polar parametrization (\ref{w-polar}), and a re-scaling of the radial coordinate $\zeta \to  \zeta/k_n$,  the above metric simplifies to
\begin{eqnarray}\label{simply-Banados}
ds^2=L^2\frac{d\zeta^2}{\zeta^2}+L^2\,k_n^2\frac{(1+\chi^2)^2}{\chi^2}\frac{dr^2}{r^2} +L^2\,k_n^2\frac{(1-\chi^2)^2}{\chi^2}d\theta^2\,,
\end{eqnarray}
where 
\begin{eqnarray} \label{chi-gen}
\chi(\zeta,r,\theta)=\frac{\zeta}{r}\left|\frac{\partial w}{\partial z}\right|\,.
\end{eqnarray}
Notice, that the later rescaling in $\zeta$ requires to do the inverse rescaling, namely $\zeta\to k_n \zeta$ near the boundary to recover the appropriate near boundary metric. In this form, it is clear that the $\chi=1$ surface must play a special role, since on those points the determinant of the metric vanishes, and the angular coordinate becomes ill defined.  This later feature happens at the end points of the intervals, thus, one might expect that the bulk continuation of this feature might signal the location of Dong's brane.
We will study this metric in detail in the next subsections with special emphasis in the $n\to 0$ limit. 
 
\subsubsection{Half-infinite interval: two dimensional Rindler\label{2dRindler}}
 It is convenient to revisit the case of half-infinite interval from the perspective of the above exact two dimensional solution. For a half-infinite interval, starting at $z=0$, the fluid velocity coincides with the angular coordinate in the $z$ plane, thus, the transformation (\ref{uni-map}) becomes the trivial identification $w(z)=z=re^{i\theta}$. Using (\ref{simply-Banados}) we obtain
\begin{eqnarray}\label{rindler-banados}
ds^2=L^2\frac{d\zeta^2}{\zeta^2}+L^2k_n^2\frac{(1+\chi^2)^2}{\chi^2}\frac{dr^2}{r^2} +L^2 k_n^2\frac{(1-\chi^2)^2}{\chi^2}d\theta^2\,,
\end{eqnarray}
where 
\begin{eqnarray}\label{chirindler}
\chi(\zeta,r,\theta)=\frac{\zeta}{r}\,.
\end{eqnarray}
In this case, the function $\chi(\zeta,r,\theta)$ is independent of $\theta$, thus, the surfaces of constant $\chi$ are cones given by $\zeta(\chi, r)=\chi r$. As previously observed, the $\chi=1$ surface is special since $\det g=0$ there. This implies that our solution should be trusted only in the region $0\leq \zeta \leq r$. On the other hand, the induced metric on the $\chi=1$ surface is 
\begin{eqnarray}\label{brane-rindler}
ds_{ind}^2=(4k_n^2+1)L^2\frac{dr^2}{r^2}=\frac{L^2}{n^2}\frac{dr^2}{r^2}\,.
\end{eqnarray}
This means that the $\chi=1$ surface is not two dimensional but actually one dimensional. Thus, there is no natural way to continue the Euclidean solution beyond the region $0\leq \zeta \leq r$.  In other words, the space-time described above is a complete Euclidean spacetime. Nevertheless this surface might be singular, and thus it deserves further scrutiny.

Now, we would like to explore the region around the $\chi=1$ surface. For that purpose let us define the radial coordinate $\ell$ through 
\begin{eqnarray}
\chi^2:=1-\ell\,, \qquad{\rm with  }\quad \ell \ll 1\quad \to \quad  \log \zeta= \log r+\frac12 \log(1-\ell)\approx  \log r-\frac{\ell}{2}\,.
\end{eqnarray}
In this form, we write a diagonal metric in terms of the radial coordinate $\ell$, in the regime  $ \ell \ll 1$
\begin{eqnarray}\label{con-def-rindler}
ds^2\approx (4k_n^2+1)L^2\frac{d\tilde{r}^2}{\tilde{r}^2}+\frac{ L^2k_n^2d\ell^2}{4k_n^2+1}+L^2k_n^2\ell^2d\theta^2\,,
\end{eqnarray}
where we have redefined the boundary radial coordinate $r$ as
\begin{eqnarray}
\frac{d\tilde{r}}{\tilde{r}}=\frac{d{r}}{{r}}-\frac{d\ell}{2(4k_n^2+1)}\,.
\end{eqnarray}
We further reescale the radial coordinate $\ell$ as
\begin{eqnarray}
\tilde{\ell}:=\frac{L\,k_n \ell}{\sqrt{4k_n^2+1}}\quad{\to}\quad ds^2\approx (4k_n^2+1)L^2\frac{d\tilde{r}^2}{\tilde{r}^2}+d\tilde{\ell}^2+(4k_n^2+1)\tilde{\ell}^2 d\theta^2\,.
\end{eqnarray}
Since the boundary angular coordinate $\theta$ has period $2\pi$, we find from the above metric, that the geometry near $\chi=1$ has a conical singularity 
\begin{eqnarray}
\Delta \theta=2\pi \(\frac{n-1}{n}\)\,,
\end{eqnarray}
in perfect agreement with Dong's brane proposal. In other words, the $\chi=1$ surface describes the location of the brane dual to $\tilde{S}_n$, and the metric (\ref{rindler-banados}) is the fully backreacted dual geometry. Notice also that the location of the brane in the holographic coordinate corresponds to $\zeta=r$ which is proportional to the local temperature of the dual boundary fluid. For Rindler space $\beta=2\pi r$.

Finally, we can use this geometry to compute the refined R\'enyi entropy for half interval using (\ref{Dong-prescription}). We need to compute the area of the brane, which in our case is one dimensional and it is given by the induced metric on $\chi=1$, which was described in (\ref{brane-rindler}). Thus, we get 
\begin{eqnarray}\label{ref-2d-rindler}
\tilde{S}_n(A)=\frac{L}{4G_N\, n}\int_\epsilon^{L_{IR}}\frac{dr}{r}=\frac{L}{4G_N\, n}\log\(\frac{L_{IR}}{\epsilon}\)\,,
\end{eqnarray}
where $\epsilon$ is a cutoff, and $L_{IR}$ is an IR regulator.
This formula gives the refined R\'enyi entropy for Rindler pace. One can easily derived from (\ref{ref-2d-rindler}) the more familiar R\'enyi entropy formula for half-infinite interval integrating \eqref{Refined-Sn-Def} from $n=n$ to $n=1$. The result is
\begin{eqnarray}\label{Sn-rindler}
S_n(A)=\frac{c}{12}\(1+\frac1n\)\log\(\frac{L_{IR}}{\epsilon}\), \quad{\rm where}\quad c=\frac{3 L}{2G_N} \,,
\end{eqnarray}
where in the last formula we written in terms of the central charge using the Brown-Henneaux relation.  In summary, we found that in two dimensional Rindler, the geometry \eqref{rindler-banados}, \eqref{chirindler} describes the expectation value of $\rho^n/\tr\rho^n$ for arbitrary values of $n$, in $0<n<1$. This follows from the fact that for an interval the constrains \eqref{p-const} fixes completely the boundary tensor to be \eqref{2d-stress-tensor} with $1/n^2\to 4k_n^2$.

\subsubsection{Arbitrary number of intervals \label{2dNIntervals}}

The analysis of the previous section generalizes almost straightforward to an arbitrary number of intervals. Our starting point is (\ref{simply-Banados}) with (\ref{chi-gen}). First, let us identify the surface $\chi=1$. In the general case we have:
\begin{eqnarray}
\zeta(r,\theta)=\(\frac{1}{r}\left|\frac{\partial w}{\partial z}\right|\)^{-1}=\frac{1}{\left|\partial_z \log(w)\right|}=\frac{(2\pi)^{-1}}{\left|\partial_z \alpha\right|}=\frac{(2\pi)^{-1}}{\left|\partial a\right|}\,,
\end{eqnarray}
where in the last equality the gradient is computed in the $\mathbb{R}^2$ geometry of the complex $z$ plane.  This shows explicitly that on $\chi=1$, $\zeta(r,\theta)$ equals the local temperature $\beta(r,\theta)/2\pi$, described in the polar coordinates of the $w$ plane.

The surface of zero determinant, however, is no longer one dimensional. If one computes its induced metric, one gets
\begin{eqnarray}\label{brane-gen}
ds_{\rm ind}^2=L^2\frac{\(\partial_r\zeta(r,\theta)dr+\partial _\theta \zeta(r,\theta)d\theta\)^2}{\zeta^2(r,\theta)}+4L^2k_n^2\frac{dr^2}{r^2} \,,
\end{eqnarray}
where $\partial _\theta \zeta(r,\theta)\neq 0$ in the most general case.  In general, depending on $n$, this metric could be extended beyond the zero determinant surface. 

However, in the $n\to 0$ limit, where the above geometry describes our physical problem of interest, we can restrict to the part of the induced metric that scales with $1/n^2$ or equivalently with $k_n^2$. See App. \ref{AppC} for a simple example in which we show that the volume inside the det$g=0$ wall is subleading for the $\tilde S_{n\to0}$ computation. Keeping only such part, we get effectively a one dimensional surface:
\begin{eqnarray}\label{brane-gen-2}
ds_{\rm ind}^2\underset{n\to 0}{\approx} 4L^2k_n^2\frac{dr^2}{r^2}\approx \frac{L^2}{n^2}\frac{dr^2}{r^2}\,.
\end{eqnarray}
In general, thus, we are forced to take the $n\to 0$ limit in the more general case, to obtain a simple description that can be matched with Dong's brane from the Ba\~nados solution. This is an important difference with respect to half interval, and it is due to the fact that the R\'enyi entropies for multiple intervals are non universal while the ones of half interval or a single interval are universal \cite{Faulkner:2013yia}.   

The Ba\~nados solution does not uncover for us the expected singularity around the location of the Brane due to the presence of the zero determinant wall. This wall shields the Brane and exclude us from accessing the near geometry of the brane itself even in the $n\to 0$ limit.\footnote{It is reasonable to expect a similar brane geometry as the one for Rindler in the multi-interval case despite the fact that these geometries do not describe the boundary stress tensor of the R\'enyi state, see App. \ref{AppC}.
} Fortunately, this solution does not precludes from understanding the effective emergent one dimensional geometry behind the brane  in the $n\to 0$ limit \eqref{brane-gen-2}, and therefore, to evaluate the refined R\'enyi entropy as  
\begin{eqnarray}
\tilde{S}_{n\to 0}(A)=\frac{L}{4G_N}\frac{1}{n}\sum_{j=1}^N \int \frac{dr_j}{r_j}=\frac{L}{4G_N}\frac{1}{n}\sum_{j=1}^N \int_{A_j} \sum_{i=1}^N\(\frac{dx_j}{x_j-l_i}+\frac{dx_j}{r_i-x_j}\)\,,
\end{eqnarray}
where in the first equality we use the fact that each $w$ plane covers only the region associated to a single boundary interval and thus, we need to sum over the $N$ copies. In the second equality, we use the relations (\ref{w-polar}) and (\ref{uni-map}) for $\theta=0$ or $y=0$ for simplicity, and to properly include the standard cutoff dependence. The final result for the refined R\'enyi entropy is 
\begin{eqnarray}\label{Sn-gen-2d}
\tilde{S}_{n\to 0}(A)=\frac{c}{3\,n}\( \sum_{i,j} \log|r_j-l_i|-\sum_{i<j} \log|r_j-r_i|-\sum_{i<j} \log|l_j-l_i|-N\log\epsilon \)\,,
\end{eqnarray}
where we have used the Brown-Henneaux relation (\ref{Sn-rindler}) to express the R\'enyi entropies in terms of the central charge. Thus, we find that the holographic R\'enyi entropies in the $n\rightarrow 0$ limit, for an arbitrary number of intervals, has the same form as the R\'enyi entropy of the free fermion \cite{Casini:2009sr}, in line with the expected universality \cite{Agon:2023tdi}. 

\section{Discussion and future perspective \label{Sec:Discussion}}

In this paper we have argued that the $n\rightarrow 0$ limit of the holographic R\'enyi entropy is related to the solution of an irrotational Euclidean perfect fluid with vortex-like boundary conditions at the boundary of the region of interest. The problem is then reduced to one of partial differential equations in the boundary. The fluid solution however, plays a role in the construction of the bulk geometry. This is similar in spirit to the real time holographic description of fluids by black branes, though the details differ as explained in Sec. \ref{Testing n0}.
To our knowledge this is the first time that an Euclidean fluid equation has found an application. In general, Euclidean time rotation of real time fluids lead to complex equations of motion due to, e.g. dissipative terms that involve a single time derivative, which obstructs a direct physical interpretations of these systems.

One could wonder about higher order corrections to our solutions away from $n\to0$. In particular, we would argue that we expect that no dissipative or any other time-reflection odd terms to appear in the boundary stress tensor since we are studying the state $\tilde \rho$ at a moment of time reflection symmetry $t=0$. This, alongside with the Einstein's equations of motion expanded order by order in $n$ can shed light on the metric corrections. 

Regarding explicit solutions to our fluid problem, we have not attempted to actually solve the fluid equations in this work. Even for the simplest case of a strip in $d\ge 3$, that corresponds to a vortex anti-vortex solution in the dimensionally reduced $d=2$ problem, this should involve numerical work. This remains an interesting subject for future investigation. 

To end, we discuss three topics connecting the results of this paper with previous literature.

\subsection{Comparison with free theories and other CFT's}

The limit $n\rightarrow 0$ was studied for free theories in \cite{Agon:2023tdi}. In this case the solution corresponds to a gas satisfying a free Boltzmann equation, where the particle densities at each point and direction is given in terms of direction dependent local temperatures. These temperatures are in correspondence with the local structure of the modular Hamiltonian \cite{Arias:2016nip,arias2017anisotropic} (see \cite{Caminiti:2025hjq} for a related development).  They are determined by Euclidean eikonal equations sourced by the boundary of the region. As in the holographic case, an entropic conserved current can be defined, but in the free case this current is the result of a local average over the particle density as a function of direction \cite{Agon:2023tdi}. 

It is then natural that in the highly interacting holographic case the Boltzmann gas gets replaced by a fluid, where the angle dependent temperatures get replaced by the unique local temperature of the fluid. It is unclear however what is the meaning of this fluid temperature in terms of the local structure of the modular Hamiltonian in this case.  

This suggests to see the free and holographic cases as two extremes regarding the strength of the interaction and naturally posses the question of the status of other CFT's. A naive conjecture would be that there should be a fluid description in the general case, somewhat intermediate between the free Boltzmann case and the perfect fluid. However, it is difficult to imagine the equations for such a fluid. The difficulty being that these should be a deformation of a perfect fluid but without dimension-full coefficients. The impossibility to conceive such a deformation prompt us to believe that it is quite possible that a perfect fluid be the description of the $n\rightarrow 0$ limit for all interacting CFT's, disregarding the strength of the interaction.\footnote{This poses no necessary inconsistency because there could be an order of limits between the limit of vanishing interaction strength and $n\rightarrow 0$.} 
It would be interesting to explore corrections in the gauge theory coupling constant $\lambda$ or the number of colors $N$ in the holographic setting. In this sense, we note that the works \cite{Galante:2013wta,Bianchi:2016xvf} found universality for spheres in the $n\rightarrow 0$ limit (in terms of the Stefan Boltzmann constant) for different gravity theories in the bulk.

\subsection{Comparison with other works on holographic R\'enyi entropy for \texorpdfstring{$n<1$}{n<1}}

Recently, a modified brane prescription for the holographic dual to the R\'enyi entropy was proposed \cite{Dong:2023bfy} in the language of a fixed area state expansion of the density matrix $\tilde \rho$, see also \cite{Penington:2024jmt}. This new proposal differs from the previous one \cite{dong2016gravity}, precisely when $0<n<1$. According to the authors, for sufficiently small $n$ the saddle configuration that reproduces the correct $\tilde S_n$ is not given by any of the saddles that the original prescription \cite{dong2016gravity} considered. In such circumstances, the prescription in \cite{Dong:2023bfy} includes geometries where topologically distinct brane configurations live simultaneously on the various topologically inequivalent saddles with tensions that add up to the tension of the original prescription $T_n=(n-1)/4nG_N$. Consider, for instance, the case studied in \cite{Dong:2023bfy} in which there are two topologically distinct saddles say $\gamma_1$ and $\gamma_2$ with areas $A_1$ and $A_2$. Whenever $n>1$ the authors show that the prescription in \cite{dong2016gravity} is enough and $\tilde S_n$ is given by the candidate saddle of minimal action, either $\gamma_1$ or $\gamma_2$. However, for small enough $n$ another transition occurs and, below some (model dependent) value $n_0<1$, $\tilde S_n$ can be computed from an Euclidean path integral involving a saddle that has two branes of topologies $\gamma_1$ and $\gamma_2$ at the same time with $A_1=A_2$ which they call ``diagonal'', where the tension is distributed as $T_{n,1}=x\,T_n$  and $T_{n,2}=(1-x)\,T_n$ for some $x \in  [0,1]$ which is determined by an extremization condition. Their results do not involve an $n\to0$ limit. 

\begin{figure}\centering
\includegraphics[width=.9\linewidth]{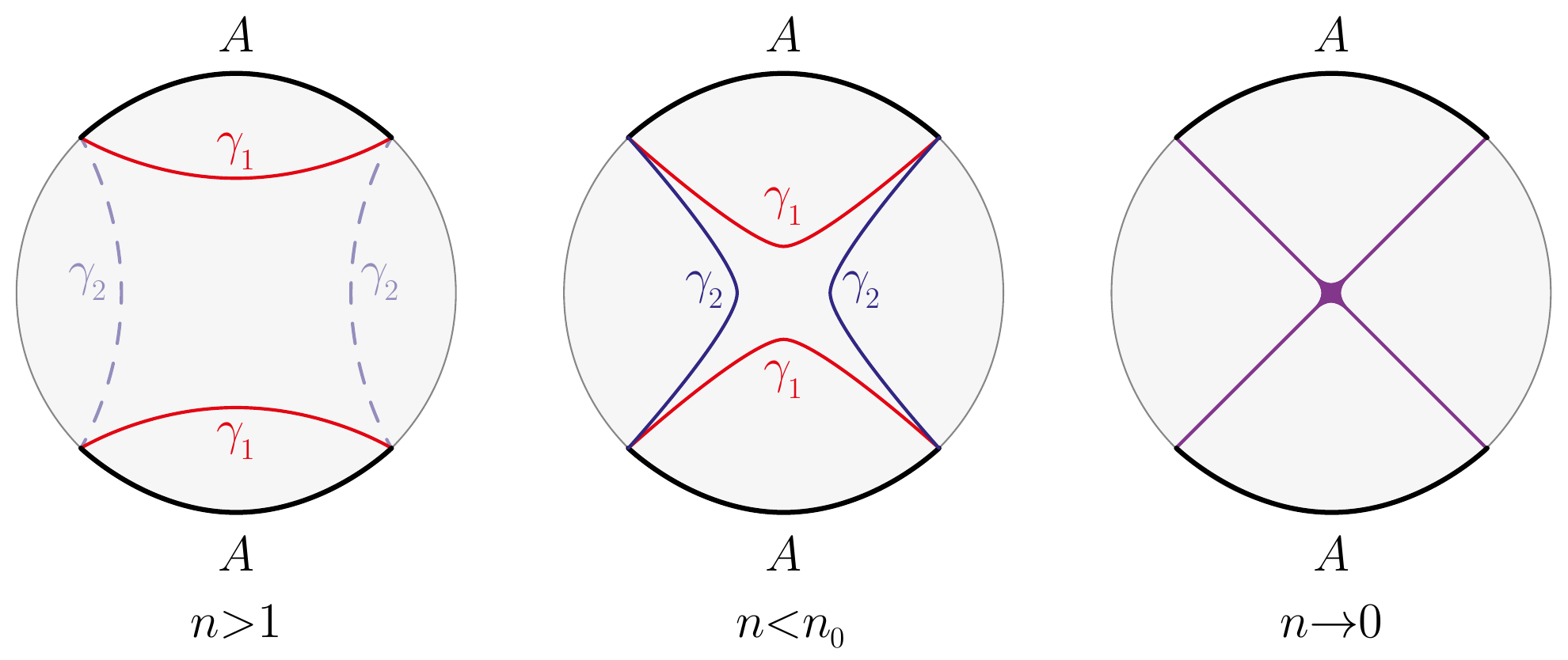} 
\caption{Brane geometry representations for different values of $n$. On the left figure, we illustrate the back-reacted geometry of a brane $\gamma_1$ in the standard  $n>1$ case. Here, $\gamma_1$ and $\gamma_2$ are the two topologically distinct branes, homologous to the entangling region $A$, with associated areas $A_1$ and $A_2$, respectively.  The holographic refined R\'enyi entropy is given by the area of $\gamma_1$ which we assumed to be minimal: $A_1<A_2$. 
 The centered figure illustrates an example of a geometry that contains both back-reacted branes $\gamma_1$ and $\gamma_2$, with $A_1=A_2$. This is the relevant geometry for sufficiently small $n$ ---below a certain model dependent $n_0<1$---  as it was recently argued in \cite{Dong:2023bfy}. 
On the right figure, we schematically draw the topology of our ansatz in which a single brane-like object carries the full tension. This geometry can be thought of as the limiting $n\to0$ case of the figure in the center.} \label{Fig1}
\end{figure}

We would like to argue that our solutions support these expectations, generalize their argument for general dimensions and any number of disconnected entangling regions and explain the mechanism behind it in the $n\to 0$ limit. Namely, for small $n$ we expect (see App. \ref{AppC}) that the bulk regions between the branes in the ``diagonal'' saddles shrink, making their associated surfaces approach each other and coalesce in the $n\to 0$ limit, at which point their areas agree. In that case, the tensions of the almost superposing branes adds up giving rise to an effective single saddle with the same tension than in the original prescription. In the two dimensional case, following the analysis from the Ba\~nados solutions (section \ref{section-2d-exact} and App. \ref{AppC}), all these effects appear including the shrinking of the bulk region inside the wall. In our general ansatz (section \ref{BlackBraneAprox}), this is implemented by the explicit construction of a single saddle with a fully connected topology. 
We illustrate this discussion in Fig. \ref{Fig1}.

\subsection{Density of states for the reduced system}
It is well known that knowledge of the R\'enyi entropies associated to the state $\rho_A$, provides information about their eigenvalue distribution. Namely, 
\bea
S_n(A)=\frac{1}{1-n}\log\tr \rho^n_A\,, \qquad{\rm where} \qquad  \tr \rho^n_A=c_n\int_0^\infty e^{-nE} \rho_A(E)
\eea
normalized such that $\tr \rho_A=1$. Thus inverting the above equation, one can find $\rho_A(E)$ as a function of $S_n(A)$ via some integral transform. Indeed, this inversion was studied recently in the context of the holographic R\'enyi entropies and a holographic formula for the density of states was derived \cite{Baiguera:2024ffx}.  In \cite{Agon:2023tdi}, it was conjectured that the R\'enyi entropies in the $n\to 0$ limit are universal and given by  
\bea\label{Renyi-0-density}
S_{n}(A)\sim \sigma_s\frac{g(A)}{n^{d-1}}\,, \quad\to \quad e^{\sigma_s\frac{g(A)}{n^{d-1}}}=\lim_{n\to 0}c_n \int_0^\infty dE e^{-nE} \rho_A(E)\,.
\eea
It is thus, intuitive that such knowledge will suffice to determine the universal behavior of the high modular energy density of states for the reduced system. To see this, instead of inverting the right hand side equation (\ref{Renyi-0-density}), we can approximate the integral by using the following asymptotic ansatz for density of states $\rho_A(E)\sim e^{k_A E^{a}}$. This results in
\bea\label{int-density}
c_n \int_0^\infty dE e^{-nE} \rho_A(E) \sim e^{(1-a)k_A\(\frac{n}{a\, k_A}\)^{\frac{a}{a-1}}}\int_0^{\infty} e^{-\frac{(E-E_*)^2}{2\sigma^2(E_*)}}dE\,,
\eea
where we evaluate the integral via the stationary phase approximation, assuming the distribution is highly peaked, $\sigma(E_*)\ll E_*$, which is true in the $n\to 0$ limit provided $0<a<1$. Comparing \eqref{int-density} with  \eqref{Renyi-0-density} one finds that $a=(d-1)/d$, and 
\bea\label{density-asympt}
\rho_A(E)\sim e^{k_A E^{\frac{d-1}{d}}}\,\quad{\rm where }\quad k_A=\frac{d \sigma^{\frac1d}_s}{(d-1)^{\frac{d-1}{d}}}g^{\frac1d}(A)\,.
\eea
This result agrees with the results presented in \cite{Baiguera:2024ffx} when the entangling region is spherical and thus $g(A)={\rm vol}_{\rm reg}({\mathbb{H}}^{d-1})/(2\pi)^{d-1}$.  Our result \eqref{density-asympt}, thus  
provides an interesting generalization for the asymptotic density of states of a reduced system which applies both for holographic and non-holographic theories and arbitrary entangling regions, along the line of the universality advocated in this work.

\section*{Acknowledgements}

We thank Alex Belin, Umut G\" ursoy, Juan Pedraza, Gonzalo Torroba and Vaios Ziogas for helpful discussions and correspondence.
CA is supported by the Netherlands Organization for Scientific Research (NWO)
under the VICI grant VI.C.202.104. 
HC was partially supported by CONICET, CNEA and Universidad Nacional de Cuyo, Argentina. 
PJM was partially supported by CONICET and UNLP. 

\appendix

\section{Euclidean perfect fluids\label{App:perfect-fluids}}
A fluid is an effective coarse grained description of a many body system which can be described in terms of a local velocity vector and some local notion of temperature. In particular the stress energy tensor of such a system will depend on those variables via constitutive relations which involve additional parameters characterizing the specific fluid. 
In a conformal field theory, the stress-energy tensor of a fluid satisfies the conservation and tracelessness condition
\begin{eqnarray}
T^{\mu}_{\,\,\,\,\, \mu}=0\,,\qquad {\rm and }\qquad \nabla^\mu  T_{\mu \nu}=0\,.
\end{eqnarray} 
In absence of any dissipative effects the appropriate description is that of a perfect fluid, which will be completely fixed by its local temperature $\beta$ and fluid velocity vector\footnote{ In relativistic fluids there is a gauge freedom in defining the ``local velocity'' which we fix by using the Landau gauge. This is, we define $u_\mu$ to be the only unitary eigenvector of $T^{\mu\nu}$ with eigenvalue $-\varepsilon$, i.e. $ T^{\mu\nu}u_\nu=-\varepsilon \,u^\mu$.} $u^\mu$
\begin{equation}\label{per-T}
T_{\mu \nu}=\frac{\sigma_{\varepsilon}}{(d-1)}\frac{1}{\beta^d}\(g_{\mu \nu}-du_\mu u_\nu\)\,,
\end{equation}
where the Stefan Boltzmann constant describes the relation between the energy density and temperature $\varepsilon=-u_\mu u_\nu T^{\mu\nu}=\sigma_{\varepsilon} \beta^{-d}$. The fluid equations of a perfect fluid follows simply from conservation of the stress-energy tensor (\ref{per-T}). It is a set of closed partial differential equations relating $u^\mu$ and $\beta$. For instance, by explicit evaluation of  $\nabla^\mu  T_{\mu \nu}=0$ one gets:
\begin{eqnarray}\label{div-T}
\nabla_\mu \log\beta-du_\mu (u\cdot \nabla)\log \beta+u_\mu (\nabla\cdot u)+(u\cdot \nabla)u_\mu=0\,.
\end{eqnarray}
Projecting this equation along the direction of $u$ one obtains a relation between the directional derivative of $\log\beta$ and the divergence of $u$, namely
\begin{eqnarray}\label{per-fluid-1}
\nabla \cdot u=(d-1) (u\cdot \nabla)\log \beta\,.
\end{eqnarray}
Plugging this relation back into (\ref{div-T}) one can express the derivatives of $\log \beta$ purely in terms of the velocity and its first derivatives as
\begin{eqnarray}\label{per-fluid-2}
\nabla_\mu \log \beta=\frac{1}{d-1}u_\mu (\nabla\cdot u)-(u\cdot \nabla) u_\mu\,.
\end{eqnarray}
In fact, the vector equation (\ref{per-fluid-2}) is equivalent to (\ref{div-T}), as one can check by projecting both  (\ref{per-fluid-2}) and  (\ref{div-T}) over every possible independent direction. This means in particular that (\ref{per-fluid-1}) follows from projecting (\ref{per-fluid-2}) along $u$.
In fact, equation (\ref{per-fluid-1}) is equivalent  to the conservation equation
\begin{eqnarray}
\nabla_\mu J^\mu=0 \qquad {\rm with } \qquad J^\mu=c_1 \frac{u^\mu}{\beta^{d-1}} \,,
\end{eqnarray}
where $c_1$ is any arbitrary constant. Thus the existence of a conserved and traceless perfect stress-energy tensor implies the existence of a conserved current. In our fluid context such current can be interpreted as the entropy current. While in a general fluid such current is not conserved, in the perfect fluid case it is. Using this interpretation we fix $c_1=d\, \sigma_{\varepsilon}/(d-1)$ , thus, we have 
\begin{eqnarray}\label{entropy-current}
J_s^\mu=\frac{d\, \sigma_{\varepsilon}}{(d-1)}\frac{u^\mu}{\beta^{d-1}}\,.
\end{eqnarray}

\subsection{Irrotational perfect fluids }\label{irrot}

One can consider perfect fluids with some extra special features. An interesting one is the irrotational or free vorticity condition. This considers the following
\begin{eqnarray}\label{vorticity}
w_\mu:=\frac{u_\mu}{\beta}, \quad {\rm obeys}\quad  \nabla_{[\mu } w_{\nu]}=0\,.
\end{eqnarray}
For the fluid equations, this has an important consequence. Recall that the fluid equations are given by (\ref{per-fluid-2}). However, it is convenient to separate that equation in two parts, one which is (\ref{per-fluid-1}) ( obtained by projecting over $u$), and the other one is obtained by plugging $(\nabla\cdot u)$ from (\ref{per-fluid-1}) into (\ref{per-fluid-2}). The result is the equation 
\begin{eqnarray}\label{per-fluid-3}
(u\cdot \nabla)\(\frac{u_\mu}{\beta}\)=\nabla_\mu \(\frac{1}{\beta}\)\,,
\end{eqnarray}
supplemented by (\ref{per-fluid-1}). Equation (\ref{per-fluid-3}) has the advantage of being independent of the number of spacetime dimensions. Importantly, a fluid which satisfies the vorticity condition (\ref{vorticity}) would automatically satisfy  (\ref{per-fluid-3}), thus such fluids will only need to satisfy (\ref{per-fluid-1}) for it to be a perfect fluid.  This is, we would only need to impose the conservation of the entropy current (\ref{entropy-current}).

The zero vorticity condition allows us to write the temperature and velocity (locally) in terms of a scalar field $\alpha$ via  
\begin{eqnarray}\label{u-beta-alpha}
\frac{u_\mu}{\beta}= \nabla_\mu \alpha\,,\qquad \beta=\frac{1}{|\nabla \alpha|}\,. 
\end{eqnarray}
The conservation equation (\ref{per-fluid-1}) now write as an ``equation of motion''
\be\label{eq-fluid-scalar}
\nabla^2 \alpha - \frac{(d-2)}{\beta}\, (\nabla \beta)(\nabla \alpha)=0\,,\quad \beta^2=\frac{1}{(\nabla \alpha)^2}\,, 
\ee
Irrotational vortexes are codimension-2 defects of the theory that serve as sources for eq. \eqref{eq-fluid-scalar} and manifests as branch points in $\alpha$. The stronger the vortex the stronger the jump at the cut. This is, in the presence of a vortex at the boundary of the entangling region $\partial A$, of strength $h$, we have
\begin{equation}\label{bc-fluid-scalar}
\oint_{\Gamma}dx^\mu \omega_\mu=\oint_{\Gamma}dx^\mu\frac{u_\mu}{\beta}=h
\end{equation}  
where $\Gamma$ is any closed oriented contour that encircles $\partial A$. For any other contour $\Gamma'$ that does not encircle $\partial A$ the integral trivially vanishes. The condition \eqref{bc-fluid-scalar} generalizes straighforwardly for any number of vortexes. It is expected that these conditions alongside the vanishing of the temperature and velocity at infinity would force a unique solution for $\alpha$. 
Notice that since \eqref{eq-fluid-scalar} is non-linear in $d>2$ finding any solution to this problem is still a formidable task even for two straight vortices.
A solution for a single straight vortex to the equations above is well known and leads to the Rindler solution. For $d=2$ the equation becomes the linear Laplace equation, and the solutions are easily obtained, as shown in section \ref{Sec:d=2}.

\subsection{Two dimensional flows}
In this section, we would like to revisit the problem of finding solutions to perfect fluids, with planar symmetry.  In other words, fluids which are effectively two dimensional. We will start by rewriting the fluid equations in terms of $u^\mu$ and $\beta$ as well as in terms of $j^\mu$ and $\beta$. As reviewed above, the perfect fluid equations (\ref{div-T}) can be separated in to two. One corresponds to the conservation of the entropy current (\ref{per-fluid-1})
\bea
j^\mu:=\frac{u^\mu}{\beta^{d-1}}\quad \to \quad \nabla_\mu j^\mu=0\quad\to \quad
\nabla \cdot u=(d-1) (u\cdot \nabla)\log \beta\,,
\eea
while (\ref{per-fluid-3}) characterizes the circulation of the fluid. In particular it is trivially satisfied for a fluid with zero circulation, namely $u_\mu=\beta \, \partial_\mu \alpha$\,. In terms of the current it looks like
\begin{eqnarray}\label{circ-free}
(u\cdot \nabla)\(\frac{u_\mu}{\beta}\)=\nabla_\mu \(\frac{1}{\beta}\)\,,\quad \to \quad \beta^{d-1}( j\cdot\nabla)\(\beta^{d-2} j_\mu\)=-\frac{1}{\beta^2}\nabla_\mu \beta\,.
\end{eqnarray}
As mentioned around (\ref{per-fluid-1}) the projection of (\ref{circ-free}) along $u$ becomes an identity. Thus, in two dimensions we can project the above equation onto the direction orthogonal to $u$, namely $\epsilon_{\mu \nu} u^\nu$, and thus obtain a scalar equation that is equivalent to (\ref{circ-free}). We can further simplify the problem by writing the current in terms of a scalar function $\phi$ as:
\bea\label{current-phi}
j_\mu=\epsilon_{\mu \nu} \nabla_\nu \phi\,, \qquad {\rm with }\qquad j^2=|\nabla \phi|^2=\frac{1}{\beta^{2(d-1)}}\,.
\eea
By doing this, the conservation equation becomes an identity while (\ref{circ-free}) projected onto $\epsilon_{\mu \nu} u^\nu=\nabla_\mu \phi/|\nabla \phi|$ gives
\bea
\beta^{2(d-1)}j_\mu j_\nu \nabla_\mu \nabla_\nu \phi-\frac{1}{\beta}\nabla_\mu\phi \nabla_\mu \beta=0\,.
\eea
Using the relation $\delta_{\mu \nu}=u_\mu u_\nu +v_\mu v_\nu$ where $u$ and $v$ are two mutually orthogonal unit vectors, and taking $u$ to be the fluid velocity while $v=\nabla\phi/|\nabla \phi|$. We get:
\bea
(\delta_{\mu \nu}-\beta^{2(d-1)}\nabla_\mu \phi \nabla_\nu \phi) \nabla_\mu \nabla_\nu \phi-\frac{1}{\beta}\nabla_\mu\phi \nabla_\mu \beta=0\,.
\eea
This expression can be further simplified to 
\bea\label{phi-beta}
\nabla^2\phi+(d-2)\frac{1}{\beta}\nabla_\mu \phi \nabla_\nu \beta=0\,,
\eea
where we used the fact that $(\nabla \phi)^2=\beta^{2(1-d)}$. Thus we are left with a single differential equation for two scalar functions which satisfy the relations (\ref{current-phi}). Notice that this fluid lives in arbitrary dimensions but has the particular feature that its dynamics involves only the coordinates on a plane. This formulation is dual to the formulation (\ref{eq-fluid-scalar}), (\ref{u-beta-alpha}), which is based on the assumption that the fluid has no circulation, which allows the introduction of the scalar function $\alpha$ and solves (\ref{circ-free}) exactly. The conservation equation thus becomes (\ref{eq-fluid-scalar}). In the present case we introduced a scalar function $\phi$ via \eqref{current-phi} to identically solve for the conservation equation \eqref{per-fluid-1} and we are left with the circulation equation \eqref{phi-beta}. Thus, for two dimensional flows the above formulation holds more generally. 

{\bf In two dimensions:} the fluid equations reduces to
\bea
j^2=|\nabla \phi|^2=\frac{1}{\beta^2}\,, \qquad \nabla^2\phi=0\,, \qquad{\rm where}\qquad j^\mu=\epsilon_{\mu \nu}\nabla_\nu \phi\,.
\eea
which can be contrasted with the dual fluid equations
\bea
|\nabla \alpha|^2=\frac{1}{\beta^2}\,,\qquad \nabla^2\alpha=0\,, \qquad{\rm where}\qquad j^\mu=\frac{u^\mu}{\beta}=\nabla^\mu \alpha\,.
\eea
The duality relations $\epsilon_{\mu \nu}\nabla_\nu \phi=\nabla^\mu \alpha$  are the Cauchy-Riemann equations indicating $\alpha$ and $\phi$ are real and imaginary parts of a single analytic function. 

\subsection{Gradient Expansion}\label{Grad-Exp}

Consider now possible corrections to the perfect conformal fluid model \eqref{per-T}. Since we are interested in holographic applications we want the fluid to remain conformal. A standard tool to incorporate these corrections is called gradient or derivative expansion. This will propose a formal expansion of the stress tensor in terms of $\beta$ and $u_\mu$ and its derivatives and in principle contains all possible terms compatible with the symmetries of the theory. We write
\begin{equation}\label{grad-T-1}
T_{\mu \nu}=\frac{\sigma_{\varepsilon}}{(d-1)}\frac{1}{\beta^d}\(g_{\mu \nu}-du_\mu u_\nu\)+\Pi^{(1)}_{\mu \nu}+\Pi^{(2)}_{\mu \nu}+\dots\,,
\end{equation}
where $\Pi^{(i)}$ are all terms containing $i$-th gradients of $\beta$ and $u_\mu$. Now, already at this level one can see that the structure of these contributions are of the form
\begin{equation}
\Pi^{(i)}_{\mu \nu}=\frac{\pi^{(i)}_{\mu \nu}}{\beta^{d-i}}\;, \qquad g^{\mu\nu}\Pi^{(i)}_{\mu\nu}=0
\end{equation}
where $\pi^{(i)}$ is an adimensional function of only the metric and fluid velocity containing only a number $i$ of derivatives and the overall $\beta$ factor fixes dimensions. The tracelessness is mandated by conformal symmetry. The fact that $\pi^{(i)}$ contains only the velocity and the metric comes from using \eqref{per-fluid-2} (valid to first order in gradients in our own expansion) to show that $\nabla_\mu \beta$ carries an extra power of $\beta$ and can be ultimately put in terms of velocity gradients. Thus, any a priori dependence in $\nabla_\mu \beta$ in our formal expansion at order $i$ can be put in terms of velocity gradients and or moved towards the $i+1$ term. Whilst the most general structure of $\pi^{(1)}$ is known and its parameters understood as dissipative effects in the fluid, a general structure for $\pi^{(i)}$ with $i>1$ and the interpretation of their free parameters is still an open problem.

Up to this point, there is no sense in which any of the terms is dominant over the other. A dominance is established if we find a proper way to disregard derivatives of the DOFs of the fluid with respect with their values at each point \cite{Hubeny:2010wp}. The characteristic size of the changes in $\beta$ and $u_\mu$ are estimated in our set-up by a characteristic length $L$ of our subsystem $A$ under study, this is
\begin{equation}
L^{-1}\sim \nabla_{\nu}u_\mu \sim \nabla_{\nu}\ln \beta
\end{equation} 
where we have again used \eqref{per-fluid-2} to relate the gradients of the velocity and temperature. This is to be compared to the only other dimensionful parameter $\beta=1/T$. So long as
\begin{equation}
 \frac{\beta}{L}\ll1
\end{equation}
our derivative expansion as presented in \eqref{grad-T-1} is a proper perturbative expansion in $\epsilon$.
We now have a clear criteria to disregard the $i+1$-th contribution with regards to the $i$-th. This is, factoring out the $\beta^{-d}$ in front of the perfect fluid contribution one finds that as long as 
\begin{equation}\label{grad-condition}
\beta \,\nabla_{\nu}u_\mu \ll 1\,, \qquad \beta\, \nabla_{\nu}\ln \beta \ll 1
\end{equation}

In our set-up, these limits are guaranteed as follows. The size $L$ of the system is fixed for the entire problem, but an overall factor of $n\to0$ is introduced in our fluid temperature $\beta_n\sim n\to0$ for our solutions. An overall small factor in front of the temperature easily realizes \eqref{grad-condition}. 

\section{Holographic R\'enyi ${n\to 0}$ entropies from the Schottky \\ uniformization method \label{Faulkners-method}}

In this Appendix, we study the $n\to 0$ R\'enyi entropies for the multi-interval geometry in holographic two dimensional CFTs from the perspective of the Schottky uniformization techniques introduced in \cite{Faulkner:2013yia}. The recipe described in \cite{Faulkner:2013yia} starts with the second order differential equation 
\bea\label{mon-psi}
\psi''(z)+\frac12 T_{zz}\psi (z)=0\,,
\eea
where $T_{zz}(z)$ is a simple rescaling of (\ref{tete}) 
\bea\label{Tzz-Faulkner-B}
T_{zz}(z)=\Delta \sum_i \(\,\frac{1}{(z-z_i)^2}+\frac{p_i}{z-z_i}\)\,.
\eea
Here $\Delta=(n^2-1)/(2n^2)$ and $z_i$ are the end points of the intervals. The $p_i$'s are accessory parameters which are determined by imposing some trivial monodromy condition on the solutions $\psi(z)$ around a set  of $N$ cycles $\gamma_N$ among the set of independent cycles defined on the complex plane $z$ with the points $z_i$ removed. The various choices of $N$-cycles give different solutions for the $p^\gamma_i$ which in turns determines the ``saddle'' R\'enyi entropies via the differential equation 
\bea\label{saddle-Renyi}
\frac{\partial S^\gamma_n}{\partial z_i}=-\frac{c (n+1)}{12 \, n}p^\gamma_i\,.
\eea
The R\'enyi entropy is one whose value is minima among the different saddles $S_n=\underset{\gamma}{\rm min} \, S_n^\gamma$.

In the $n\to0$ limit $\Delta\sim -1/(2n^2)$ and the $p_i$'s can be at most ${\cal O}(1)$ in that limit. Thus, in this regime the stress tensor in \eqref{mon-psi} is or order $1/n^2$ which allows an asymptotic analysis of the associated differential equation at leading order in $1/n$. The following ansatz for the solutions
\bea
\psi(z)=e^{A(z)}, \qquad{\rm leads\,\, to}\qquad \psi''(z)=e^{A(z)}\[(A'(z))^2+A''(z)\]\,,
\eea
where $A\sim 1/n$ can cancel the stress tensor term in the differential equation. The term $A''(z)$ is order $1/n$ and therefore, can be ignored within this approximation. This is known as the Eikonal approximation and the solution is trivially given by 
 \bea\label{A-phase}
A(z)=\pm \frac{1}{2 n}\int^z_{z_0} dz \sqrt{\sum_{i}\(\,\frac{1}{(z-z_i)^2}+\frac{p_i}{z-z_i}\)}\,.
\eea
In general, the varios functions $\psi(z)$ which represents different saddles, are obtained by finding the different sets of $p_i$'s that makes $\psi(z)$ to have trivial monodromies along $N$ paths that encloses an even number of points $z_i$.  Such $p_i$'s should satisfy additionally \eqref{p-const}. 

However, given the simple form of $A(z)$ in $\eqref{A-phase}$, $A(z)$ will have trivial monodromy on a circle at infinity if the argument of the square root in \eqref{A-phase} is a perfect square. Otherwise, the function $A(z)$ would not be analytic on a circle that encloses all the interval endpoints $z_i$'s.
This is the same as the perfect fluid condition argued in \ref{stress-tensor-1} on physical grounds. Let us write \eqref{Tzz-Faulkner-B} as
\bea\label{Tzz-B-2}
T_{zz}(z)=\Delta \beta(z)^2\, \qquad{\rm with}\qquad  \beta(z)=\sum_i \frac{ b_i}{z-z_i}\,.
\eea
In this case $A(z)=\beta(z)/(2n)$. Matching \eqref{Tzz-B-2} with \eqref{Tzz-Faulkner-B} results in 
\bea
\sum_i \frac{ b^2_i}{(z-z_i)^2}+\sum_{i \neq j}\frac{ b_i b_j}{(z-z_i)(z-z_j)}=\sum_i \frac{1}{(z-z_i)^2}+\frac{p_i}{z-z_i}\,.
\eea
The second sum on the left hand side of the above equation corresponds to single pole terms. Thus, after some algebra, we obtain the constrains
\bea
b_i^2=1\,, \qquad{\rm and}\qquad p_i=\sum_{j\neq i}\frac{ 2 b_i b_j}{z_i-z_j}\,, 
\eea
which implies $b_i=\pm1$. So far the sign of the $b_i$ is undetermined. However, from the fluid perspective it is easy to argue that the signs associated to the endpoints of a given interval must be opposite, thus $b_{2i}=\pm 1$ and $b_{2i+1}=-b_{2i}$. Similarly, the signs of all the left/right poles should also agree. This fixes the stress tensor completely. The overall sign will affect $\beta(z)$ but not $T(z)$. Therefore, we conclude that
\bea\label{pi-zero}
p_i=\sum_{j\neq i}\frac{ 2 (-)^{i+j}}{z_i-z_j} \,,
\eea
which fixes uniquely the auxiliary parameters of the solution. It is easy to show that such parameters satisfy the constrains \eqref{p-const}.  

In order to check the associated monodromy conditions, we plug the $p_i$'s into \eqref{A-phase} which leads to 
\bea
A(z)=\pm \frac{1}{2n}\int^z_{z_0} dz \sum_{i=1}^{2N} \frac{ (-)^i}{z-x_i}=\pm \frac{1}{2 n} \sum_{i=1}^{N}\[\log\(\frac{z-r_i}{z_0-r_i}\)-\log\(\frac{z-l_i}{z_0-l_i}\)\]\,,
\eea
where $r_i$ and $l_i$ represents right and left endpoints of the $i$ th interval. The solution has a simple monodromy condition. It has trivial monodromy on any path that encloses an even number of endpoints $z_i$ with zero total polarity (equal number of left and right endpoints).  

Given the $p_i$'s one can calculate the associated R\'enyi entropy via \eqref{saddle-Renyi}. The solution to \eqref{saddle-Renyi} that follows by partial integration of \eqref{pi-zero} gives \begin{eqnarray}
S_n=\frac{c}{6\,n}\!\!\( \sum^N_{i,j} \log|r_j-l_i|-\sum^N_{i<j} \log|r_j-r_i|-\sum^N_{i<j} \log|l_j-l_i|-N\log\epsilon \).
\end{eqnarray}
In perfect agreement with our main two dimensional result\footnote{Recall, in the $n\to0$ limit $\tilde{S}_n=d S_n$.} \eqref{Sn-gen-2d-2}.

\section{A vanishing volume inside the \texorpdfstring{det$g=0$}{detg=0} wall}\label{AppC}

In this appendix, we provide evidence that as $n\to0$, the volume inside the codim-1 det$g=0$ wall in the Bañados AdS$_3$ solution is subleading with respect to the volume outside. We do so in two different coordinate foliations of the Rindler problem, were an exact to solution is available to compare to. In particular, this supports our claim around eqs. \eqref{brane-gen}- \eqref{brane-gen-2} that the refined Renyi entropy can be computed from the Bañados metric using the area of the degenerate codim-2 surface that the det$g=0$ wall collapses to in the $n\to0$ limit, eq. \eqref{brane-gen-2}. This is also evidence in favor of our argument in Fig. \ref{Fig1} that as $n\to0$ any particular brane configuration that lives inside the det$g=0$ wall collapses to the topology of our solution. Whilst not a complete proof, we expect the results provided here to extend in a similar fashion to any number of intervals.

We begin by computing the refined Renyi entropy of the Rindler geometry from the bulk volume (disregarding the brane term) rather than the codim-2 brane area \cite{dong2016gravity}. This will allow us to compare more easily with the cases where a codim-1 det$g=0$ wall is present. We will use the Rindler solution in Bañados coordinates in the form \eqref{eq:Banados} with
\begin{equation}
{\cal L}=\frac{k_n^2}{z^2}\qquad\qquad \overline{{\cal L}}=\frac{k_n^2}{\bar z^2} \qquad\qquad z=r e^{i\theta}	\qquad\qquad \bar z=r e^{-i\theta}	
\end{equation}
which amounts to
\begin{eqnarray}\label{rindler-banados-2}
ds^2 = 
L^2\frac{d\zeta^2 }{ \zeta ^2}+L^2\frac{\left(r^2+\zeta ^2  k_n^2\right)^2 }{\zeta ^2 r^4} dr^2+L^2\frac{ \left(r^2-\zeta ^2 k_n^2\right)^2 }{\zeta ^2 r^2}d\theta^2\,,\qquad \sqrt{|g|}=\frac{L^3 \left(r^4-\zeta ^4
   k_n^4\right)}{\zeta ^3 r^3}\,,
\end{eqnarray}
The regularized coordinates domain is $\theta\sim\theta+2\pi$, $r\in[\epsilon,L_{IR}]$ and $\zeta \in [\epsilon , r/k_n)$ where $\epsilon$ and $L_{IR}$ are UV and IR cutoffs of the volume, and $\epsilon\ll L_{IR}$. As explained in the main text, the $\zeta=r/k_n$ is for this solution both the det$g=0$ surface and the location of the brane, i.e. these coordinates cover the full manifold.

The bare volume can be computed straightforwardly, i.e.
\begin{equation}
V_n=\int d^{2+1}x \sqrt{|g|}=L^3\frac{\pi}{2}\frac{ L_{IR}^2}{\epsilon
   ^2}-2\pi L^3 k_n^2 \log \left(\frac{L_{IR}}{\epsilon }\right)
\end{equation} 
where we have dropped subleading and cut-off dependent terms. To regularize it we choose the $n=1$ ($k_n=0$) volume that is the pure AdS solution. We have, for holographic 2d CFTs, \cite{dong2016gravity}
\begin{equation}\label{ref-2d-rindler-2}
\tilde{S}_n(A)=\frac{n^2}{4\pi L^2 G_N}\partial_n(V_n-V_1)
=\frac{L}{4G_N\, n}\log\left(\frac{L_{IR}}{\epsilon}\right)\,,
\end{equation} 
in exact agreement with \eqref{ref-2d-rindler}.

We now can arrive to the bulk solution for the configuration of a single interval from two different points of view. First and simplest, we can recognize that such a bulk solution is connected to that of Rindler and obtain it from a conformal map as in \cite{Casini:2011kv} landing in a set of coordinates that cover the full manifold. The Renyi entropy in the interval now has contributions from both boundaries of the interval and yields twice the result in eq. \eqref{ref-2d-rindler-2}. From a different point of view, we can land on the same manifold from a Bañados-like solution by using a stress tensor containing two twist operators and try to recover the same result from there. We will find that, in the Bañados patch, a codim-1 det$g=0$ surface arises where the Bañados patch breaks down. We will compute the volume outside the wall and check that as $n\to0$ we recover the correct result. This is equivalent in this example to show that the volume inside the wall is sub-leading with respect to the outside to estimate $\tilde{S}_n(A)$ in the $n\to0$ limit.

To obtain the Bañados solution for an interval of length $2a$, we use \eqref{eq:Banados} again but this time with 
\begin{equation}
{\cal L}=k_n^2\left(\frac{1}{a-z}+\frac{1}{z+a}\right)^2 \qquad\qquad \overline{{\cal L}}=k_n^2\left(\frac{1}{a-\bar z}+\frac{1}{\bar z+a}\right)^2
\end{equation}
The explicit form of the metric is not particularly enlightening, but one can compute the determinant of the Bañados metric,  $z=x+i y$ $\bar z=x-i y$, 
\begin{equation}
\sqrt{|g|} =\frac{L^3}{\zeta^3} \left(1-\frac{16 a^4
   \zeta^4  k_n^4}{\left(a^4+2 a^2
   \left(y^2-x^2\right)+\left(x^2+y^2\right)^2\right)^2}\right)
\end{equation}
to find that the surface
\begin{equation}
\zeta=\zeta_c=\frac{\sqrt{a^4+2 a^2
   \left(y^2-x^2\right)+\left(x^2+y^2\right)^2}}{2 a k_n} 
\end{equation}
makes $\text{det}g=0$ and it is a codim-1 surface. We then take the symmetries of the problem to compute the volume of the manifold outside of this wall as follows. First, we use the reflection symmetries in both $\{x,y\}$ to rewrite the integration as four times that of the first quadrant. On this quadrant we change variables to polar coordinates as $x=a+r\cos\theta$ and $y=r \sin\theta$ and split the integrand as
\begin{align}\label{Vn-2d-interval}
\tilde V_n=\int d^{2+1}x \sqrt{|g|}&=4\left(\int_{0}^{\frac\pi2} d\theta \int_{\epsilon}^{L_{IR}} r dr + \int^{\pi}_{\frac\pi2+\epsilon} d\theta \int_{\epsilon}^{-\frac{a}{\cos\theta}} r dr\right)\int_\epsilon^{\zeta_c}\sqrt{|g|}\nonumber\\
&=4 \left(\frac{1}{2} \pi  k_n^2 L^3 \log
   (\epsilon )+\frac{1}{2} \pi  k_n^2 L^3 \log
   (\epsilon )\right)+h(k_n)+\tilde V_1\nonumber\\
&=-2\left(2 \pi  k_n^2 L^3 \log\left(\frac{a}{\epsilon} \right)\right)+h(k_n)+\tilde V_1
\end{align}
where the integration limit $-\frac{a}{\cos\theta}$ in the radial $r$ integral on the second term is just the $x=0$ integration limit of the first quadrant and the angle is integrated only in $\theta\in[0,\pi]$ for the same reason. We have isolated the $n=1$ volume as $\tilde V_1$ and labelled a number subleading contributions in $k_n$ terms as $h(k_n)$.

From our result in \eqref{Vn-2d-interval}, it is immediate to recover the correct result for the $\tilde S_n$. The contributions coming from $h(k_n)$ to the R\'enyi entropy are either subleading as the cut-off $\epsilon$ is taken to be the smallest parameter in the problem, or are subleading in the $n\to0$ limit itself. This concludes our example.

\bibliography{EE}{}
\bibliographystyle{utphys}

\end{document}